\documentclass[11pt,letterpaper]{article}

\usepackage[utf8]{inputenc}
\usepackage[T1]{fontenc}
\usepackage{lmodern}
\usepackage{amsmath,amssymb,amsfonts}
\usepackage{graphicx}
\usepackage{booktabs}
\usepackage{hyperref}
\hypersetup{
  colorlinks=true, linkcolor=black, citecolor=black, urlcolor=blue, pdfborder={0 0 0},
  pdftitle={SPL: Orchestrating Workflows with Declarative Deterministic--Probabilistic Composition},
  pdfauthor={Wen G. Gong}
}
\usepackage{url}
\usepackage{listings}
\usepackage{xcolor}
\usepackage{longtable}
\usepackage{array}
\usepackage{calc}
\usepackage{nicefrac}
\usepackage{microtype}
\usepackage[normalem]{ulem}
\usepackage{geometry}
\usepackage[section]{placeins}  

\definecolor{splkeyword}{RGB}{0,0,180}
\definecolor{splstring}{RGB}{163,21,21}
\definecolor{splcomment}{RGB}{0,128,0}
\definecolor{codebg}{RGB}{248,248,248}

\lstdefinelanguage{SPL}{
  keywords={WORKFLOW, PROCEDURE, INPUT, OUTPUT, GENERATE, INTO, CALL, PARALLEL,
            EVALUATE, WHEN, THEN, ELSE, END, WHILE, DO, ASSERT, OTHERWISE,
            SOLVE, COMMIT, LOGGING, IMPORT, EXCEPTION, EXCEPTION_TYPE,
            CREATE, FUNCTION, RETURNS, PROMPT, WITH, STATUS, DEFAULT,
            TYPE, TEXT, LIST, SET, INT, FLOAT, BOOL, IMAGE, AUDIO, VIDEO},
  keywordstyle=\color{splkeyword}\bfseries,
  sensitive=false,
  morecomment=[l]{--},
  commentstyle=\color{splcomment}\itshape,
  morestring=[b]",
  stringstyle=\color{splstring},
  basicstyle=\ttfamily\small,
  breaklines=true,
  showstringspaces=false,
  tabsize=4,
  frame=single,
  backgroundcolor=\color{codebg},
  captionpos=b,
}

\providecommand{\tightlist}{\setlength{\itemsep}{0pt}\setlength{\parskip}{0pt}}
\title{Orchestrating Workflows with Declarative\\[4pt]
  Deterministic--Probabilistic Composition}

\author{Wen G. Gong\\
  \small \texttt{wen.g.gong@gmail.com}\\
  \small Independent Researcher}

\date{}

\begin{document}

\maketitle

\begin{abstract}
We present SPL (Structured Prompt Language), a declarative language that composes \emph{deterministic} and \emph{probabilistic} computation modes in a single specification. While existing frameworks separate these---orchestration systems (AutoGen, CrewAI, LangGraph) for LLM calls, symbolic tools (SymPy, SageMath, Lean) for computation---SPL unifies them. It provides $\texttt{GENERATE}/\texttt{EVALUATE}$ for probabilistic computation and $\texttt{SOLVE}/\texttt{ASSERT}$ for deterministic computation, sharing syntax, variable bindings, and runtime routing. A \texttt{.spl} specification runs unchanged across local nodes (Ollama), cloud APIs (OpenRouter, Anthropic), and distributed grids (Momagrid), with model and verifier selection deferred to invocation time.

We validate SPL through an extensive $78$-recipe cookbook and a controlled $1{,}200$-run experiment ($10 \text{ models} \times 20 \text{ problems} \times 2 \text{ arms} \times 3 \text{ repetitions}$; the $20$ problems span $6$ difficulty tiers). The solver arm achieves $82$--$93\%$ machine-verified correctness (sonnet-4-6: $85\%$, gemma4:e2b: $93\%$) while the LLM-only arm measures output production without mathematical verification, making the comparison one of \emph{verified correctness} against \emph{unverified fluency}. A backend difficulty gradient emerges (SymPy $78\%$, Sage $54\%$), and the dominant failure mode is $\texttt{solver\_error}$ (kernel-rejected expressions), not format non-compliance.
\end{abstract}


\section{Introduction}\label{introduction}

\subsection{The Problem: Fragmentation in LLM Programming}\label{the-problem-fragmentation-in-llm-programming}

Building LLM-powered workflow systems today demands a daunting intersection of skills. A practitioner must master prompt engineering for effective LLM interaction, Python programming for orchestration logic, API wrangling across disparate model providers, and manual state management for multi-step reasoning. Each popular framework --- LangGraph, AutoGen {[}1{]}, CrewAI {[}2{]} --- introduces its own abstractions, API surfaces, and execution models, creating a fragmented ecosystem where knowledge transfers poorly and vendor lock-in is the norm.

This situation echoes a pattern the data industry has seen before. In the early days of relational databases, programmers wrote imperative C and COBOL code to traverse data structures, manage cursors, and handle errors --- all in application-specific ways. The introduction of SQL in the 1970s transformed this landscape by providing a declarative specification layer: users described \emph{what} data they wanted, not \emph{how} to retrieve it. This separation enabled decades of optimizer innovation underneath a stable surface language. The current state of LLM programming is analogous to the pre-SQL era: powerful capabilities buried under layers of imperative glue code.

The barrier to entry compounds the problem. To build even a simple self-refining workflow --- one that drafts, critiques, and revises its own output --- a developer must write 80--150 lines of Python across multiple framework abstractions. This excludes the vast population of domain experts, analysts, and researchers who understand their workflows but lack the software engineering skills to express them in imperative orchestration code.

\subsection{The Two-Mode Gap}\label{the-two-mode-gap}

All existing frameworks --- declarative and imperative alike --- operate in a single computation mode. LLM-centric systems (LangGraph, DSPy, AutoGen) produce approximate, non-reproducible outputs shaped by the model's training distribution. Symbolic tools (SymPy, SageMath, Lean) produce exact, reproducible, machine-verifiable results. Neither world speaks the other's language within a unified programming model. We call this the \textbf{two-mode gap}.

Real agentic tasks routinely require both. A homework assistant must plan a solution in natural language (probabilistic), execute each algebraic step exactly (deterministic), verify the result (deterministic), then explain it to the student (probabilistic). The coupling between modes may be \emph{loose} --- distinct phases passing data once --- or \emph{tight}: an \texttt{ASSERT} failure immediately triggers a corrective \texttt{GENERATE}, which revises the plan and re-enters the symbolic engine. Both patterns arise in practice; both require application-specific glue code to coordinate state transfer and error propagation at every mode crossing.

The gap is not an engineering inconvenience --- it is an expressiveness ceiling. No existing workflow language lets the programmer declare ``compute this step exactly'' and ``generate this step approximately'' in the same specification. SPL closes it through \emph{declarative composition}: the programmer specifies what each step computes and which mode it inhabits; the runtime resolves the boundary, manages state transfer, and enforces verification gates --- without glue code.

The dual-process theory of human cognition {[}21{]} offers a useful analogy --- but requires a critical correction. Kahneman's framing names the two systems \emph{fast} and \emph{slow}, conflating epistemic mode with cognitive speed. That conflation misleads in a computational setting: discovering quantum mechanics was intuitive, exploratory, and extraordinarily slow; once Schrödinger's equation was established, \emph{solving} it became deterministic and mechanical --- slow by hand, fast by computer. Speed tracks the implementation, not the mode.

The relevant distinction is epistemic, not temporal. System 1 is \emph{probabilistic and intuitive}: it pattern-matches, generates hypotheses, and produces approximate outputs shaped by prior experience --- regardless of wall-clock time. System 2 is \emph{deterministic and mechanical}: it applies explicit rules to produce exact, reproducible results --- regardless of whether those rules execute in milliseconds or minutes. In SPL's mapping: \texttt{GENERATE} and \texttt{EVALUATE} are System 1 primitives (LLM inference, approximate, quality-assessed); \texttt{SOLVE} and \texttt{ASSERT} are System 2 primitives (symbolic kernel, exact, machine-verifiable). A SymPy kernel may compute a derivative faster than the LLM does; a Lean 4 proof may take longer. Neither fact changes which mode each step inhabits. Mode --- not speed --- is the organizing principle.

\subsection{Our Contributions}\label{our-contributions}

We present SPL, a declarative language that composes both computation modes in a single coherent specification. The primary contributions of this paper are:

\textbf{C1. Declarative two-mode composition.} SPL provides \texttt{SOLVE} and \texttt{ASSERT} --- primitives that route computation to a live Python kernel (IPython) rather than an LLM --- alongside \texttt{GENERATE} and \texttt{EVALUATE} for the probabilistic mode. A single declarative specification freely composes both modes; variable bindings flow between them through a shared \texttt{@variable} namespace without programmer-managed marshaling. The mode boundary is visible in the source and managed entirely by the runtime.

\textbf{C2. The DODA principle.} Design Once, Deploy Anywhere: a single \texttt{.spl} specification is the invariant. The choice of model provider (Ollama, OpenRouter, Anthropic, Momagrid) and verifier (SymPy, SageMath, Lean) is resolved at invocation time via CLI flags, not embedded in the specification. The same file runs on a developer's laptop and on a distributed inference grid without modification --- declarative composition is the mechanism that makes this possible.

\textbf{C3. The verifier ladder.} We implement a three-rung hierarchy of symbolic verifiers --- SymPy (algebraic), SageMath (mathematical), Lean 4 (formal proof) --- as ascending correctness guarantees composable into any SPL workflow via \texttt{ASSERT}. R1 and R2 are evaluated empirically in the 1200-run experiment; R3 (Lean 4) is implemented with a working bridge (15/15 unit tests passing) and demonstrated as a design contribution --- full experimental evaluation of R3 is future work. Each rung is a Python-callable tool registered at runtime; the workflow specification is agnostic to which rung is active. The ladder is itself declared, not coded: the escalation logic is \texttt{ASSERT} chains in \texttt{.spl}, not Python conditionals.

\textbf{C4. Controlled empirical evaluation.} A 1200-run experiment (10 models $\times$ 20 problems $\times$ 2 arms $\times$ 3 repetitions; problems span 6 difficulty tiers) on symbolic mathematics provides quantitative evidence of the two-mode composition benefit and surfaces a structured capability hierarchy: models that score 100\% in the probabilistic arm (LLM-only math) may score 0\% in the two-mode arm (solver), not because they cannot solve the math but because they cannot produce the structured decomposition the deterministic engine requires. This reveals that format-compliance for declarative composition is a separable capability from mathematical reasoning.

\subsection{Paper Organization}\label{paper-organization}

Section 2 surveys related work, positioning SPL against imperative frameworks, declarative systems, and neurosymbolic AI. Section 3 presents the SPL language design --- the full primitive set, semantics, and formal grammar. Section 4 formalizes the verifier ladder. Section 5 describes the implementation: how the two-mode executor dispatches probabilistic nodes to the adapter layer and deterministic nodes to the kernel layer, with model provider and verifier both selected at invocation time. Section 6 presents the 1200-run empirical evaluation (r=3). Section 7 discusses strengths, limitations, and future directions. Section 8 concludes.

\section{Related Work}\label{related-work}

\subsection{The Evolution of Data Programming}\label{the-evolution-of-data-programming}

The history of data programming offers a clear precedent for the transition SPL proposes. In the 1970s, IBM's System R project showed that a declarative query language (SQL) could match or exceed the performance of hand-coded navigational database access, because the declarative form exposed optimization opportunities invisible in imperative code {[}3{]}. By the 1980s, SQL had become the industry standard, and Oracle's PL/SQL extension showed that procedural control flow (loops, variables, exception handling) could be layered on top of declarative queries without sacrificing the optimizer's ability to reason about individual \texttt{SELECT} statements {[}4{]}.

The parallel to LLM programming is direct. Today's imperative frameworks --- LangGraph's state graphs, AutoGen's conversation loops, CrewAI's role-based agents --- are the equivalent of pre-SQL navigational code: powerful, flexible, and impossibly fragmented. SPL proposes that the same declarative revolution can occur for LLM orchestration, following the same evolutionary arc: atomic declarative queries (\texttt{PROMPT}) compose into procedural workflows (\texttt{WORKFLOW}) that an optimizer can eventually rewrite and route without altering the specification.

A critical difference from the SQL era: data queries are deterministic. LLM orchestration is not. This demands new primitives --- \texttt{EVALUATE} for semantic branching, \texttt{WHILE} for quality-gated iteration, \texttt{EXCEPTION} for probabilistic failure modes --- that have no SQL equivalent. And it demands a second computation mode entirely: \texttt{SOLVE} and \texttt{ASSERT} for the cases where a result must be exact rather than approximate.

\subsection{Imperative Orchestration Frameworks}\label{imperative-orchestration-frameworks}

\textbf{AutoGen} {[}1{]} focuses on multi-agent conversation patterns where agents exchange messages. Its strength is natural dialogue flows; its limitation is that orchestration logic is embedded in Python and tightly coupled to the runtime.

\textbf{CrewAI} {[}2{]} introduces a role-based metaphor where specialized agents collaborate. Intuitive for team-like workflows; Python-first with no path to compilation or optimization.

\textbf{PydanticAI} takes a code-first approach, prioritizing developer ergonomics and type safety. Like the above, it operates exclusively in the probabilistic mode with no symbolic integration layer.

None of these frameworks provides a language-level mechanism to say ``execute this step deterministically and verify the result before the workflow proceeds.''

\subsection{Declarative LLM Systems}\label{declarative-llm-systems}

\textbf{LMQL} {[}5{]} (ETH Zurich, 2022) pioneered SQL-like LLM interaction with output constraints (type, length, regex), achieving 26--85\% cost reduction through constrained decoding. LMQL operates at the query level --- individual prompts with output validation --- and does not provide constructs for multi-step workflows or deterministic integration.

\textbf{DSPy} {[}6{]} (Stanford, 2023) introduces declarative modules (\texttt{Predict}, \texttt{ChainOfThought}) with a compiler that automatically optimizes prompts via few-shot bootstrapping. DSPy's ``declarative'' is at the module-interface level: it optimizes which prompts to use, not how to orchestrate multi-step workflows. It provides no iteration, no exception handling, no symbolic integration, and no formal grammar.

\textbf{SGLang} {[}7{]} (UC Berkeley, 2023) provides structured generation with runtime optimizations (RadixAttention, up to 6.4$\times$ throughput). SGLang optimizes inference execution below the level of agentic patterns; it does not address workflow orchestration.

\textbf{Agent Spec} {[}8{]} (Oracle Labs, 2025) defines a YAML/JSON schema for framework-agnostic agent specification. A configuration format, not an executable language: it lacks control flow, iteration, and exception handling within the specification itself.

\textbf{eBay DSL} {[}9{]} (December 2025) separates workflow specification from implementation across multiple backend languages, achieving 60\% reduction in development time. Focused on enterprise pipeline configuration; does not formalize LLM-specific primitives or symbolic integration.

\textbf{Agentics 2.0} {[}10{]} (IBM, March 2026) introduces a logical transduction algebra where LLM calls are typed semantic transformations composable via algebraically grounded operators. The closest concurrent work in formalism, but algebraic rather than SQL-inspired, and still operating exclusively in the probabilistic mode.

\textbf{Compiled AI} {[}11{]} (Trooskens et al., April 2026) addresses non-determinism through compile-time LLM invocation: the language model runs once at specification time to generate a deterministic executable artifact; no LLM is invoked at runtime. This achieves strong execution guarantees at the cost of losing all runtime adaptability --- no \texttt{EVALUATE} branching, no \texttt{WHILE} quality gates, no \texttt{EXCEPTION} recovery. SPL takes the opposite stance: both modes remain live and equally accessible at runtime, interleaved through a single declarative specification. The tradeoff is explicit: SPL retains full probabilistic expressiveness (multi-shot LLM calls, adaptive routing, exception handling) alongside deterministic verification.

\textbf{Blueprint First} {[}12{]} (Qiu et al., August 2025) proposes an engineering pattern that decouples workflow logic from the generative model. Expert-defined procedures are codified into an ``Execution Blueprint'' (deterministic engine), and the LLM is invoked only for bounded sub-tasks within it. While this echoes SPL's DODA principle --- the blueprint as invariant, the LLM as pluggable component --- the approach remains a design pattern rather than an executable language. Blueprint First offers no symbolic verification layer (no \texttt{SOLVE}/\texttt{ASSERT} equivalent), no formal grammar, and no shared variable namespace allowing seamless data flow between modes. The blueprint is a manually maintained engineering artifact; SPL's specification is the executable, compiled form.

\subsection{LLM+Solver Composition}\label{llmsolver-composition}

A distinct line of work explores coupling language models with symbolic solvers via mode composition: the LLM translates a natural language problem into a solver encoding, the solver executes deterministically, and the LLM translates the result back to natural language.

\textbf{LLM+P} {[}13{]} (Liu et al., April 2023) pioneered this pattern: language models generate PDDL planning encodings; a classical optimal planner (Fast Downward) solves them; the LLM narrates the solution. This foundational work established the ``two-phase'' pattern --- probabilistic translation, deterministic execution, probabilistic interpretation --- that has influenced subsequent systems. The limitation: the LLM is restricted to a translator role between fixed input/output schemas (NL $\leftrightarrow$ PDDL); there is no multi-step workflow composition, shared variable namespace, or first-class language for orchestrating the two phases.

\textbf{MCP-Solver} {[}14{]} (Szeider et al., January 2025; SAT 2025) wraps constraint solvers (MiniZinc, PySAT, Z3, Clingo/ASP) as Model Context Protocol tools. The LLM builds a solver encoding through conversational tool calls, the solver executes, and the LLM interprets results. Like LLM+P, this uses a \emph{tool-call pattern}: the LLM invokes the solver as a black box via a protocol-defined schema. There is no formal workflow language, no explicit mode boundary visible in source, and the variable-passing contract is implicit in tool schemas rather than explicit in a grammar.

\textbf{DUPLEX} {[}15{]} (Hua et al., March 2026) applies dual-system composition to robotic task planning: a lightweight ``fast'' system maps entity relations to PDDL via LLM; a classical planner executes; a ``slow'' system (heavier LLM with solver diagnostics) activates on failure. DUPLEX exhibits the PLAN → SOLVE → ASSERT → EXPLAIN lifecycle that SPL formalizes as a language pattern. However, DUPLEX is confined to the PDDL domain; modes do not share variables --- they communicate through rigid schema --- and the composition is hard-coded to the two-phase structure, not visible as a declarative grammar.

Across all three systems, the LLM is a component \emph{within} a larger orchestration pattern, not a co-equal participant with explicit mode boundaries visible in source. SPL differs fundamentally: \texttt{SOLVE} and \texttt{ASSERT} are language primitives (not tool calls), the mode boundary is visible in the grammar, and variables flow between modes through a unified \texttt{@variable} namespace without schema marshaling. In a tool-call system, the developer writes explicit serialization/deserialization logic in Python to pass an LLM's string output to a SymPy tool and parse the result back; in SPL, the \texttt{\{@variable\}} interpolation syntax makes this a language-level feature managed by the runtime.

\subsection{Neurosymbolic AI}\label{neurosymbolic-ai}

A growing body of work combines neural and symbolic computation, but at the model or algorithm level rather than the workflow language level.

\textbf{AlphaGeometry} {[}16{]} (DeepMind, 2024) solves Olympiad geometry problems by alternating a neural language model (generating proof steps) with a symbolic deduction engine (verifying them). The neuro-symbolic loop is hard-coded to the geometry domain and implemented as a custom Python pipeline. SPL generalizes this pattern to arbitrary domains: the \texttt{SOLVE}/\texttt{ASSERT} primitives express any such loop declaratively, and the verifier (geometry engine, SymPy, Lean) is a pluggable runtime choice rather than a hard-coded dependency.

\textbf{Neural theorem proving} systems (PACT, LeanDojo {[}17{]}, HyperTree Proof Search {[}18{]}) use LLMs to generate Lean or Coq proof tactics, with the proof assistant as the verifier. SPL's verifier ladder places Lean at rung R3 --- the highest correctness guarantee --- accessible from any workflow via \texttt{ASSERT}.

\textbf{Program synthesis} systems (Codex {[}19{]}, AlphaCode {[}20{]}) generate code and execute it to verify functional correctness. SPL's \texttt{SOLVE} construct is the declarative equivalent of this loop: the programmer specifies the expression template; the kernel executes it; \texttt{ASSERT} verifies the result --- without the programmer writing the execution and verification harness in Python.

Recent work on composable neuro-symbolic architectures offers insights into the two-mode approach. \textbf{Compositional AI Beyond LLMs} {[}23{]} (ASPLOS 2026) examines from a systems/hardware perspective the characteristics of neuro-symbolic-probabilistic architectures, finding distinct memory and compute profiles and consistent performance gains over monolithic LLMs of comparable size. \textbf{Symbolic Seams} {[}24{]} (Schuler et al., March 2026) proposes ``symbolic seams'' --- explicit architectural breakpoints with typed boundary objects --- as a design principle for composable systems, conceptually aligned with the mode boundary that SPL makes visible at the language level.

The common thread across these systems is that neuro-symbolic integration is implemented as bespoke application logic rather than as a language primitive. SPL's contribution is to make that integration a first-class, reusable construct expressible in few keywords.

\subsection{Positioning Summary}\label{positioning-summary}

\textbf{Table 1: System Comparison}

\begin{longtable}[]{@{}
  >{\raggedright\arraybackslash}p{(\columnwidth - 10\tabcolsep) * \real{0.1667}}
  >{\raggedright\arraybackslash}p{(\columnwidth - 10\tabcolsep) * \real{0.1667}}
  >{\raggedright\arraybackslash}p{(\columnwidth - 10\tabcolsep) * \real{0.1667}}
  >{\raggedright\arraybackslash}p{(\columnwidth - 10\tabcolsep) * \real{0.1667}}
  >{\raggedright\arraybackslash}p{(\columnwidth - 10\tabcolsep) * \real{0.1667}}
  >{\raggedright\arraybackslash}p{(\columnwidth - 10\tabcolsep) * \real{0.1667}}@{}}
\toprule\noalign{}
\begin{minipage}[b]{\linewidth}\raggedright
System
\end{minipage} & \begin{minipage}[b]{\linewidth}\raggedright
Paradigm
\end{minipage} & \begin{minipage}[b]{\linewidth}\raggedright
Grammar
\end{minipage} & \begin{minipage}[b]{\linewidth}\raggedright
Semantic Eval
\end{minipage} & \begin{minipage}[b]{\linewidth}\raggedright
Both modes @runtime
\end{minipage} & \begin{minipage}[b]{\linewidth}\raggedright
Sym. Integration
\end{minipage} \\
\midrule\noalign{}
\endhead
\bottomrule\noalign{}
\endlastfoot
LangGraph & Imperative & No & No & No & Manual (Python) \\
AutoGen & Imperative & No & No & No & Manual (Python) \\
CrewAI & Imperative & No & No & No & Manual (Python) \\
LMQL & Declarative & Yes & No & No & No \\
DSPy & Declarative & No & No & No & No \\
SGLang & Declarative & Yes & No & No & No \\
LLM+P & Two-phase & No & No & No & Partial (PDDL) \\
Blueprint First & Declarative & No & No & No & No \\
Compiled AI & Declarative & No & No & No & No \\
MCP-Solver & Tool-call & No & No & Yes (ad hoc) & Via protocol \\
DUPLEX & Two-phase & No & No & No & Yes (PDDL domain) \\
Agent Spec & Declarative & Yes (schema) & No & No & No \\
eBay DSL & Declarative & Yes & No & Yes & No \\
Agentics 2.0 & Algebraic & Yes (algebra) & No & No & No \\
AlphaGeometry & Neuro-sym. & No & No & Yes (hard-coded) & Yes (domain-specific) \\
\textbf{SPL} & \textbf{Declarative} & \textbf{Yes (EBNF)} & \textbf{Yes} & \textbf{Yes (first-class)} & \textbf{Yes (pluggable)} \\
\end{longtable}

SPL is the only system where (1) both computation modes have first-class status in the formal grammar, (2) the mode boundary is visible in source via \texttt{SOLVE}/\texttt{ASSERT}, (3) variables are shared through a unified \texttt{@variable} namespace without tool schemas, and (4) adaptive control flow (\texttt{EVALUATE}, \texttt{WHILE}, \texttt{EXCEPTION}) operates across both modes at runtime. The critical distinction from every other declarative system is that SPL's declarativity extends across the mode boundary: the \texttt{.spl} source specifies composition of both modes; no Python glue bridges them.

\section{The SPL Language}\label{the-spl-language}

\subsection{Design Principles}\label{design-principles}

SPL is governed by the following design principles:

\textbf{P1. Declarative composition over imperative implementation.} The programmer specifies \emph{what} each step computes and \emph{which mode} it inhabits --- not which model, which verifier, or which execution path achieves it. The two modes may be woven loosely (distinct probabilistic and deterministic phases passing data between them) or tightly (fine-grained interleaving where \texttt{ASSERT} failure immediately re-enters \texttt{GENERATE}); SPL handles both ends of this spectrum with the same primitives. The runtime resolves the mode boundary; the \texttt{.spl} source is the invariant across both coupling styles, both model providers, and both deployment environments.

\textbf{P2. Two-mode primitives as co-equal first-class constructs.} SPL provides primitives for both computation modes --- System 1 (probabilistic, intuitive) and System 2 (deterministic, mechanical) in the dual-process sense {[}21{]}, reinterpreted epistemically rather than temporally (see \S\ref{the-two-mode-gap}). The probabilistic mode (\texttt{GENERATE}, \texttt{EVALUATE}, \texttt{WHILE}, \texttt{EXCEPTION}) addresses the characteristics of LLM computation: approximate, non-reproducible, quality-assessed, cost-sensitive. The deterministic mode (\texttt{SOLVE}, \texttt{ASSERT}) addresses the characteristics of symbolic computation: exact, reproducible, machine-verifiable. Neither mode is subordinate to the other in the language design --- their composition in a single specification is the defining property of SPL.

\textbf{P3. Shared variable space.} Outputs of \texttt{GENERATE} steps are available as inputs to \texttt{SOLVE} steps, and vice versa, through a single \texttt{@variable} namespace. The programmer does not marshal values between modes; the runtime manages the boundary. Two syntactic forms govern variable use: \texttt{@name~TYPE} is the \emph{declaration/binding} form (used in \texttt{INPUT}/\texttt{OUTPUT} signatures, \texttt{SOLVE} targets, and assignment), while \texttt{\{@name\}} is the \emph{interpolation/evaluation} form (used inside \texttt{PROMPT} templates, \texttt{SOLVE} expressions, and \texttt{ASSERT} predicates). The executor substitutes each \texttt{\{@name\}} reference with the current value of \texttt{@name} before dispatch. This is the mode-crossing seam, and it is the same mechanism as SPL f-strings (\texttt{f"..."}) applied uniformly across all string contexts.

\textbf{P4. DODA --- Design Once, Deploy Anywhere.} The \texttt{.spl} file is the invariant. Physical decisions --- which LLM, which verifier, which infrastructure --- are resolved at invocation time via \texttt{-\/-adapter}, \texttt{-\/-model}, and \texttt{-\/-kernel} flags. The same specification is the logical description of the workflow regardless of where it runs.

\textbf{P5. Compilation target.} The language is designed to produce an optimizable intermediate representation (IR), enabling future optimizer passes (GENERATE merging, SELECT caching, model routing) without changing the surface language.

\subsection{The SQL Analogy}\label{the-sql-analogy}

The mental model of SPL is grounded in its parallel to the SQL and PL/SQL stack:

\begin{itemize}
\tightlist
\item
  \textbf{\texttt{SELECT}} assembles the context that flows \emph{into} the LLM --- analogous to \texttt{SELECT\ …\ FROM\ …\ WHERE} gathering rows from a table.
\item
  \textbf{\texttt{GENERATE}} invokes the LLM and captures what flows \emph{out} --- analogous to the result set returned by a SQL query.
\item
  \textbf{\texttt{PROMPT}} binds \texttt{SELECT} and \texttt{GENERATE} into a named, independently executable unit --- analogous to a single SQL query.
\item
  \textbf{\texttt{WORKFLOW}} adds variables, loops, branching, and exception handling around \texttt{PROMPT} calls --- analogous to PL/SQL layering procedural control flow over SQL queries.
\item
  \textbf{\texttt{SOLVE}} routes a Python expression to the deterministic kernel --- an extension that has no SQL analogue.
\end{itemize}

This last point marks the boundary of the SQL analogy. SPL must go further because LLM orchestration requires a second computation mode that relational data access does not. A second boundary deserves acknowledgment: SQL optimization rests on relational algebra equivalences over deterministic set operations, whereas \texttt{GENERATE} samples from a probability distribution. Future SPL optimizations (batching, caching, model routing) must therefore operate at the \emph{workflow} level --- reordering independent nodes, fusing adjacent calls --- rather than applying algebraic rewrites to individual \texttt{GENERATE} invocations.

\subsection{Deterministic Primitives}\label{deterministic-primitives}

\subsubsection{SOLVE --- Kernel-Routed Computation}\label{solve-kernel-routed-computation}

\texttt{SOLVE} dispatches a Python expression to the live IPython kernel and assigns the result to a workflow variable:

\begin{verbatim}
SOLVE @derivative SYMPY := "diff({@expression}, x)"
\end{verbatim}

The \texttt{\{@variable\}} syntax binds any \texttt{@variable} currently in scope into the expression before dispatch. This is the boundary-crossing mechanism: the LLM's output (held in \texttt{@expression}) becomes the input to the symbolic engine without programmer-managed marshaling. The result (\texttt{@derivative}) is available immediately to subsequent \texttt{GENERATE} or \texttt{SOLVE} steps. The substitution is performed by the executor at the AST level --- the runtime resolves each \texttt{\{@identifier\}} reference against the variable store before constructing the kernel expression, rather than interpolating into a raw string. The kernel itself executes in a sandboxed IPython subprocess with no access to the host filesystem or network; malformed expressions raise a \texttt{SyntaxError} or \texttt{NameError} caught by the \texttt{EXCEPTION} handler, never arbitrary code execution on the host.

\texttt{SOLVE} is mode-explicit by design. The programmer cannot accidentally invoke the symbolic engine from a \texttt{GENERATE} call or the LLM from a \texttt{SOLVE} call. The mode boundary is visible in the source.

\subsubsection{ASSERT --- Verification Gate}\label{assert-verification-gate}

\texttt{ASSERT} evaluates a Boolean expression in the kernel and gates workflow continuation on the result:

\begin{verbatim}
ASSERT simplify({@sympy_result} - diff({@expression}, x)) == 0
  OTHERWISE RETURN @explanation WITH status = 'verification_failed'
\end{verbatim}

On success, the workflow continues --- the claim is machine-verified. On failure, the \texttt{OTHERWISE} body executes, typically committing a result with an error status or escalating to a higher verifier rung. \texttt{ASSERT} is the mechanism by which SPL workflows achieve correctness guarantees: a step that passes \texttt{ASSERT} is not merely plausible but proven correct by the kernel. Because \texttt{ASSERT} accepts any Python Boolean expression, tolerance-based assertions for numerical computation are naturally supported (e.g., \texttt{ASSERT abs(\{@result\} - \{@expected\}) < 1e-6}); the language imposes no restriction to exact symbolic equality.

The deterministic primitives are not limited to mathematics. Any Python-evaluable predicate works as an \texttt{ASSERT} condition:

\begin{verbatim}
-- Schema validation: verify LLM output conforms to expected JSON structure
ASSERT json.loads({@api_response}) and validate(json.loads({@api_response}), schema)
  OTHERWISE RETURN @result WITH status = 'schema_violation'

-- Length constraint: ensure generated content meets minimum requirements
ASSERT len({@report_body}) >= 500
  OTHERWISE RETRY

-- Graph property: verify generated network satisfies connectivity invariant
ASSERT nx.is_connected(G) and nx.number_of_nodes(G) == {@expected_nodes}
  OTHERWISE RETURN @graph WITH status = 'graph_invalid'
\end{verbatim}

These examples illustrate that \texttt{SOLVE}/\texttt{ASSERT} is a general \emph{kernel-routed computation mode}, not a math-specific feature. Any domain with a Python-callable verifier --- JSON schema validation, unit checking (pint), constraint solving (Z3), statistical testing (scipy) --- plugs into the same primitives. We label this mode ``deterministic'' throughout the paper to emphasize its reproducibility and machine-verifiability relative to LLM inference. Strictly, the Python kernel can execute non-deterministic code (e.g., MCMC sampling, randomized algorithms); in such cases the \texttt{ASSERT} gate still enforces a verifiable postcondition on the result, preserving the mode's role as the verification boundary regardless of whether the kernel computation itself is stochastic.

\subsubsection{CALL PARALLEL --- Concurrent Branch Dispatch}\label{call-parallel-concurrent-branch-dispatch}

\begin{verbatim}
CALL PARALLEL
  CALL arm_solver(@problem) INTO @result_solver
  CALL arm_llm_only(@problem) INTO @result_llm
END
\end{verbatim}

Branches execute concurrently via \texttt{asyncio.gather} on a single node, or route to distinct Momagrid worker nodes on a distributed grid. This construct makes A/B experiments --- deterministic arm vs.~probabilistic arm --- a language-level primitive rather than a test-harness concern. When parallel branches contain \texttt{SOLVE} or \texttt{ASSERT} steps, each branch operates on an isolated copy of the kernel namespace to prevent race conditions; results are merged into the parent scope only upon branch completion via the \texttt{INTO} binding.

\subsection{Probabilistic Primitives}\label{probabilistic-primitives}

\subsubsection{SELECT and GENERATE --- The Atomic I/O Contract}\label{select-and-generate-the-atomic-io-contract}

\texttt{SELECT} assembles the prompt from multiple context sources; \texttt{GENERATE} invokes the LLM and captures the result:

\begin{verbatim}
WORKFLOW explain_concept
  INPUT: @concept TEXT, @audience TEXT DEFAULT 'undergraduate'
  OUTPUT: @explanation TEXT
DO
  GENERATE explanation(concept, audience) INTO @explanation
    SELECT
      system_role('You are a clear, precise instructor.'),
      @concept AS concept,
      @audience AS audience
    PROMPT "Explain {@concept} to a {@audience} student in 3 sentences."
END
\end{verbatim}

\subsubsection{EVALUATE --- Semantic Branching}\label{evaluate-semantic-branching}

\texttt{EVALUATE} uses the LLM as a runtime judge for conditions that cannot be expressed as deterministic comparisons:

\begin{verbatim}
EVALUATE @explanation
  WHEN contains('undefined') OR contains('undefined symbol') THEN
    @explanation := @explanation + '\n\n[Note: terms defined in appendix]'
  WHEN = 'satisfactory' THEN
    @status := 'complete'
  ELSE
    @status := 'needs_revision'
END
\end{verbatim}

Quoted string conditions are dispatched to the LLM judge; equality comparisons (\texttt{=\ \textquotesingle{}satisfactory\textquotesingle{}}) are evaluated deterministically. This hybrid detection rule means the programmer does not annotate which branches are semantic and which are deterministic --- the parser infers it from syntax.

\subsubsection{WHILE --- Quality-Gated Iteration}\label{while-quality-gated-iteration}

\begin{verbatim}
WHILE 'explanation is incomplete or unclear' DO
  GENERATE refine_explanation(@explanation, @feedback) INTO @explanation
END
\end{verbatim}

The termination condition is evaluated by the LLM on each iteration: the executor assembles all current \texttt{@variable} values as context and asks the LLM judge whether the condition still holds. A built-in maximum iteration guard (configurable, default 5) prevents infinite loops.

\subsubsection{EXCEPTION --- LLM-Aware Error Handling}\label{exception-llm-aware-error-handling}

\begin{verbatim}
EXCEPTION
  WHEN HallucinationDetected THEN
    RETURN @result WITH status = 'hallucination_detected'
  WHEN ContextLengthExceeded THEN
    @problem := summarize(@problem)
    RETRY
  WHEN BudgetExceeded THEN
    RETURN @partial_result WITH status = 'budget_exceeded'
END
\end{verbatim}

The exception taxonomy formalizes failure modes specific to LLM computation --- hallucination, refusal, context overflow, budget violation --- that have no equivalent in conventional exception hierarchies. Detection mechanisms vary by type: \texttt{ContextLengthExceeded} and \texttt{BudgetExceeded} are raised deterministically by the adapter when API limits are hit; \texttt{ModelRefused} is detected by pattern matching on refusal templates in the response. \texttt{HallucinationDetected} is currently a proposed exception type --- the runtime does not yet implement automatic hallucination detection; the programmer can raise it explicitly via an \texttt{EVALUATE} guard that dispatches to an LLM judge. The taxonomy is designed to be extensible as detection capabilities mature.

\subsection{The Two-Mode Composition Pattern}\label{the-two-mode-composition-pattern}

Real agentic tasks follow a recurring four-stage lifecycle: \textbf{PLAN} (System 1 decomposes the problem), \textbf{SOLVE} (System 2 executes it exactly), \textbf{ASSERT} (System 2 verifies the result), \textbf{EXPLAIN} (System 1 narrates it in natural language). The following example shows this lifecycle expressed declaratively in the following SPL workflow:

\begin{verbatim}
WORKFLOW solve_with_verification
  INPUT: @problem TEXT
  OUTPUT: @explanation TEXT
DO
  -- [PLAN — System 1 / Probabilistic] LLM decomposes the problem
  GENERATE decomposition(@problem) INTO @steps_text
    SELECT @problem AS problem
    PROMPT 'Decompose this problem into symbolic computation steps,
            one per line in the format: expression|operation'

  -- [PLAN — System 1 / Probabilistic] Sanity gate before kernel handoff
  GENERATE plan_check(@steps_text, @problem) INTO @plan_verdict
    SELECT @steps_text AS plan, @problem AS problem
    PROMPT 'Does this plan correctly address the problem? Reply: pass or fail'

  EVALUATE @plan_verdict
    WHEN = 'fail' THEN
      RETURN @explanation WITH status = 'plan_sanity_error'
  END

  -- [SOLVE — System 2 / Deterministic] Kernel executes each symbolic step exactly
  CALL solve_chain(@steps_text) INTO @verified_result

  -- [ASSERT — System 2 / Deterministic] Verify the chain answer against the problem
  ASSERT verify({@verified_result}, {@problem})
    OTHERWISE RETURN @explanation WITH status = 'verification_failed'

  -- [EXPLAIN — System 1 / Probabilistic] LLM narrates the verified result
  GENERATE explanation(@verified_result, @problem) INTO @explanation
    SELECT @verified_result AS result, @problem AS problem
    PROMPT "Explain this verified solution in clear prose: {@result}"

  RETURN @explanation WITH status = 'complete'
END
\end{verbatim}

This 30-line workflow composes five mode transitions declaratively: System 1 decomposition → System 1 sanity gate → System 2 execution → System 2 verification → System 1 narration. Each transition is expressed in SPL syntax; no Python bridges the modes.

The \texttt{\{@var\}} template syntax is the \emph{formalization boundary} of SPL: the seam at which System 1 outputs (unstructured text from \texttt{GENERATE}) are cast into typed expressions consumed by System 2 (\texttt{SOLVE}/\texttt{ASSERT}). The programmer does not write serialization code; the runtime resolves variable bindings before kernel dispatch. The mode crossing is visible at the source level --- every \texttt{\{@...\}} reference marks a System 1 → System 2 handoff --- and is managed entirely by the language.

The programmer writes the same \texttt{.spl} file regardless of which LLM and which symbolic engine execute at runtime --- the declarative specification is the invariant, and the DODA principle holds across both the model axis and the verifier axis.

\subsection{Formal Grammar}\label{formal-grammar}

The SPL grammar defines the following key production rules for two-mode integration:

\begin{verbatim}
solve_stmt    ::= 'SOLVE' '@'identifier ['SYMPY' | 'SAGE' | 'LEAN'] ':=' string_expr
assert_stmt   ::= 'ASSERT' python_expr ['OTHERWISE' stmt_block]
parallel_stmt ::= 'CALL' 'PARALLEL' (call_stmt)+ 'END'
stmt          ::= ... | solve_stmt | assert_stmt | parallel_stmt
\end{verbatim}

The \texttt{SYMPY\ \textbar{}\ SAGE\ \textbar{}\ LEAN} type annotation on \texttt{SOLVE} is optional; when omitted, the runtime selects the kernel registered at invocation time. The full grammar is provided in Appendix A.

\section{The Verifier Ladder}\label{the-verifier-ladder}

\subsection{Motivation: Ascending Correctness Guarantees}\label{motivation-ascending-correctness-guarantees}

LLM-generated mathematical reasoning is probabilistic by nature: a model scoring 95\% on a benchmark is wrong on 1 in 20 problems, with no signal to the caller about which ones. Symbolic verification closes this gap --- but symbolic verification itself is not monolithic. A SymPy check of polynomial differentiation takes milliseconds and requires no external tooling. A formal Lean 4 proof of the same claim takes seconds, requires Mathlib, and produces a machine-checkable certificate that any Lean installation can re-verify independently. The two checks are not interchangeable: one is appropriate for routine computation, the other for publication-quality assertions.

The SPL verifier ladder makes this trade-off explicit and programmable. Three rungs address different cost/confidence operating points:

\begin{center}
\begin{tabular}{@{}p{0.5cm}p{2.6cm}p{3.8cm}p{3.5cm}p{3.0cm}@{}}
\toprule
Rung & Engine & What it certifies & Setup & Typical use \\
\midrule
R1 & SymPy (\texttt{python3} kernel) & Algebraic identity, calculus, matrix operations & \texttt{pip install sympy} & Routine STEM computation \\[4pt]
R2 & SageMath (\texttt{sagemath} kernel) & Number theory, geometry, polynomial rings, combinatorics & \texttt{pip install 'spl-llm[sage]'} & Deeper mathematical claims \\[4pt]
R3 & Lean 4 (\texttt{lean\_bridge}) & Machine-checked formal proof against Mathlib & \texttt{bash setup\_lean.sh -{}-with-mathlib} & Publication-quality assertions \\
\bottomrule
\end{tabular}
\end{center}

A workflow selects a rung via the \texttt{-\/-kernel} and \texttt{-\/-param\ backend=} flags at invocation time, optionally probing multiple rungs in parallel via \texttt{CALL\ PARALLEL} to find the highest verifier rung the expression satisfies. The \texttt{.spl} source is identical across all three rungs --- the DODA principle applies to verifier selection just as it applies to model selection.

\subsection{Rung R1: SymPy}\label{rung-r1-sympy}

The SymPy rung runs inside an IPython kernel started with the standard \texttt{python3} kernelspec. Seventeen mathematical operations are registered as SPL tools via \texttt{CREATE\ TOOL\_API}:

\begin{verbatim}
CREATE TOOL_API solve_step_with_sympy(expression, operation)
  '''
  import sympy as sp
  from sympy.abc import x, y, z, t, n
  ops = {
    "diff":        lambda e: sp.diff(e, x),
    "integrate":   lambda e: sp.integrate(e, x),
    "solve":       lambda e: sp.solve(e, x),
    "eigenvalues": lambda e: sp.Matrix(eval(e)).eigenvals(),
    "dsolve":      lambda e: sp.dsolve(sp.sympify(e)),
    ...   -- 12 further operations
  }
  result = ops[operation](sp.sympify(expression))
  return f"{result}|{sp.latex(result)}"
  '''
\end{verbatim}

The tool returns a \texttt{bare\_result\textbar{}human\_readable} protocol: the bare result is available as a string for subsequent \texttt{\{@var\}} interpolation; the human-readable side is written to the chain trace log. On failure the sentinel \texttt{solver\_error\textbar{}...} is returned, which the workflow detects via \texttt{EVALUATE} and exits with an explicit status rather than propagating a bad value downstream.

\subsection{Rung R2: SageMath}\label{rung-r2-sagemath}

SageMath extends SymPy's algebraic reach into number theory (Galois groups, elliptic curves, modular arithmetic), differential geometry, and combinatorics. The runtime discovers the SageMath Jupyter kernelspec automatically and raises a clear install error if absent (see Appendix C for setup). The same \texttt{SOLVE} / \texttt{ASSERT} primitives and \texttt{\{@var\}} interpolation work unchanged --- the only runtime difference is which kernelspec launches. This is the DODA principle applied at the verifier level: the workflow programmer writes identical SPL regardless of which mathematical engine executes beneath.

\subsection{Rung R3: Lean 4}\label{rung-r3-lean-4}

The Lean rung addresses a qualitatively different class of claim: not ``does this algebraic identity hold for generic symbols'' but ``is this mathematical statement a theorem provable from first principles within Mathlib.'' SPL integrates Lean 4 via \texttt{spl3/lean\_bridge.py}, which wraps the Lean REPL in a persistent session. The rung follows a three-stage protocol, each stage expressed in pure SPL: (1) \textbf{Formalize} --- the LLM translates the natural-language claim into a Lean 4 statement; (2) \textbf{Typecheck} --- the kernel verifies it compiles against Mathlib, with an LLM-driven repair loop; (3) \textbf{Prove} --- the LLM writes tactic proof code, which the kernel checks and repairs. The full SPL protocol is given in Appendix G.

The Lean rung produces a \emph{badge} --- \texttt{machine\_proved}, \texttt{statement\_checked}, \texttt{unfaithful}, or \texttt{unverified} --- which becomes the workflow's output status. A failed proof never blocks delivery: the claim is returned with its statement-checked grade, and the higher badge is withheld. This allows the Lean rung to run speculatively in a \texttt{CALL\ PARALLEL} branch while the primary answer is already delivered via the SymPy arm.

The cost/confidence trade-off across rungs is explicit: a SymPy algebraic check completes in milliseconds with zero external dependencies; a SageMath verification takes seconds and requires a separate conda environment; a Lean 4 formal proof may take tens of seconds to minutes (including Mathlib compilation on first use) and requires the Elan toolchain plus \textasciitilde2 GB of Mathlib. The DODA principle lets the workflow author choose the appropriate operating point at invocation time --- routine computation uses R1, publication-quality claims escalate to R3 --- without modifying the \texttt{.spl} source.

\subsection{\texorpdfstring{\texttt{ASSERT} as the Inter-Rung Gate}{ASSERT as the Inter-Rung Gate}}\label{assert-as-the-inter-rung-gate}

\texttt{ASSERT} is the mechanism by which the verifier ladder enforces correctness within a workflow. Each rung produces a value that can be checked, and \texttt{ASSERT} failure triggers escalation:

\begin{verbatim}
-- Try R1 first (inside a workflow body)
SOLVE @derivative SYMPY := "diff({@expression}, x)"
ASSERT simplify({@derivative} - diff({@expression}, x)) == 0
  OTHERWISE
    -- R1 could not verify; escalate to R2
    SOLVE @derivative SAGE := "SR({@expression}).diff(x)"
    ASSERT bool({@derivative} == diff({@expression}, x))
      OTHERWISE RETURN @explanation WITH status = 'r2_failed'
\end{verbatim}

This escalation pattern is idiomatic SPL declarative composition: try the cheapest rung, gate on \texttt{ASSERT}, compose with a higher rung on failure. No Python glue code is required; the entire escalation logic --- including the mode transitions between LLM narration and kernel verification --- is expressed declaratively in the \texttt{.spl} source.

\subsection{Pluggability}\label{pluggability}

The verifier ladder is open-ended: any Python-callable verifier plugs in without modifying the language or runtime. A workflow author registers a new tool via \texttt{CREATE\ TOOL\_API}, selects it via a parameter or \texttt{-\/-kernel} flag, and writes \texttt{ASSERT} conditions against its output. Beyond the three rungs demonstrated in this paper, the pattern generalizes to: unit verification (pint), graph property checking (NetworkX), constraint solving (Z3), type checking (mypy), or statistical hypothesis testing (scipy.stats). The language makes no assumption about what the kernel computes --- only that it is Python-callable and that the result is a string the workflow can inspect.

\section{Implementation}\label{implementation}

The executor parses a \texttt{.spl} source into an AST and dispatches each node to one of two engines: probabilistic nodes (\texttt{GENERATE}, \texttt{EVALUATE}, \texttt{WHILE}, \texttt{EXCEPTION}) route to the adapter layer, which normalizes calls across 14 LLM providers behind a two-method interface (\texttt{generate} / \texttt{generate\_multimodal}). Deterministic nodes (\texttt{SOLVE}, \texttt{ASSERT}) route to a persistent kernel session --- an out-of-process Jupyter kernel for symbolic computation, or a lightweight in-process substrate for tool execution. The kernel path activates only when \texttt{SOLVE} or \texttt{ASSERT} is encountered; otherwise no kernel is started. Adapter and kernel are both selected at invocation time (\texttt{-\/-adapter}, \texttt{-\/-model}, \texttt{-\/-kernel}) and never named in the \texttt{.spl} source --- the deterministic mode of DODA mirrors the probabilistic one.

\begin{figure}
\centering
\includegraphics[width=\textwidth]{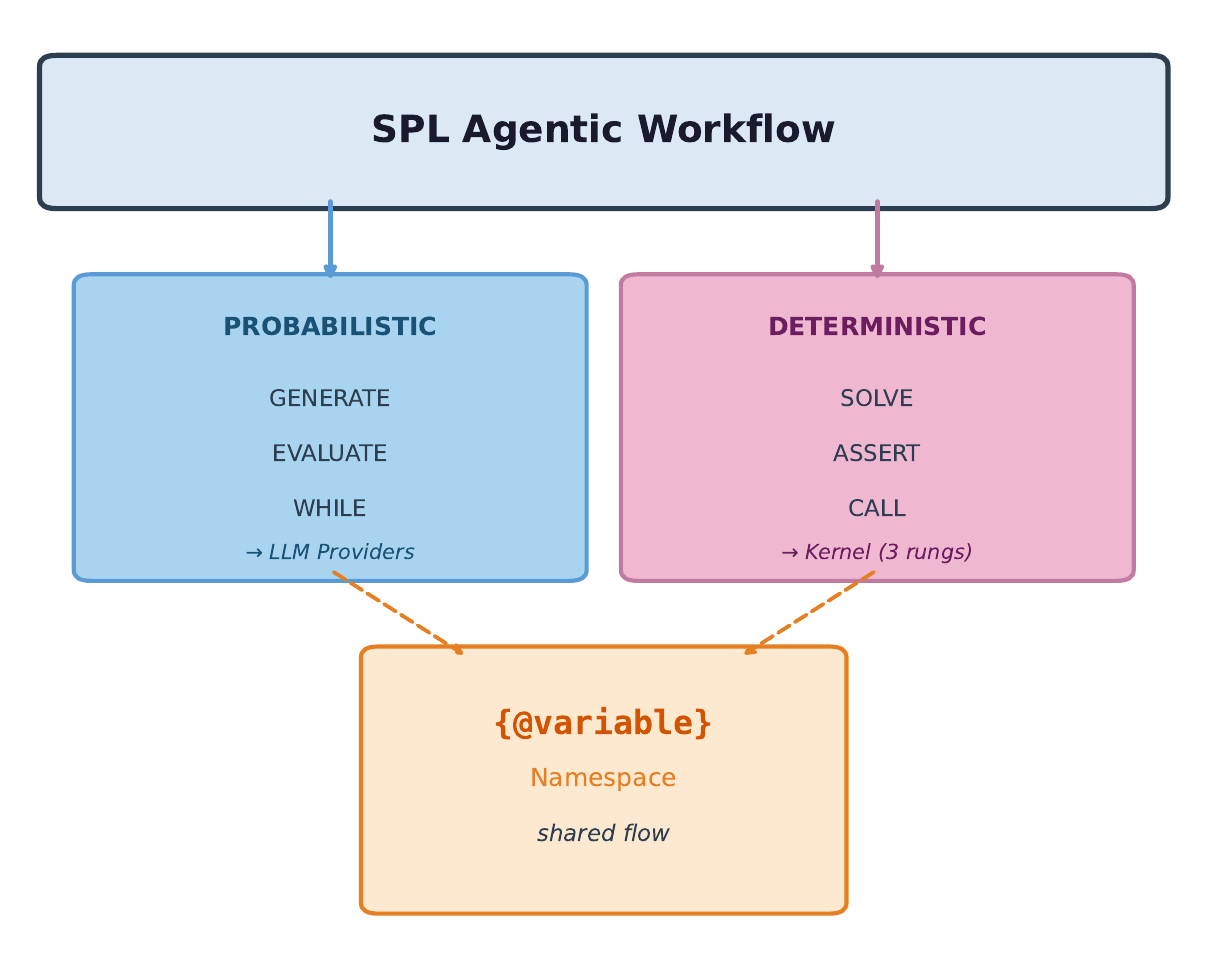}
\caption{Execution Pipeline: The executor dispatches probabilistic nodes to the adapter layer (14 LLM providers) and deterministic nodes to a pluggable kernel, with both modes sharing state through the \{@variable\} namespace.}
\end{figure}

The execution pipeline routes \texttt{GENERATE}/\texttt{EVALUATE} to the adapter layer and \texttt{SOLVE}/\texttt{ASSERT} to the kernel, with both modes sharing state for seamless mode crossing without marshaling code.

The two modes share state through \texttt{\{@var\}} interpolation: before a \texttt{SOLVE} expression is dispatched to the kernel, the executor inlines current SPL variable values into the expression string. The result is captured and bound back to a \texttt{@var}, making it available to subsequent \texttt{GENERATE} or \texttt{EVALUATE} nodes. This variable-passing contract is the integration seam between the two modes; no explicit marshalling code is required in the workflow.

The declarative form is not merely syntactic convenience --- it enables optimizations that are impossible in imperative orchestration code. Because the executor sees the full AST before dispatch, it can fuse adjacent \texttt{GENERATE} calls into a single batched API request, cache \texttt{SELECT} results across workflows, reorder independent \texttt{SOLVE} steps for maximum kernel throughput, and route cheap sub-tasks to smaller models --- all without modifying the \texttt{.spl} source. An imperative Python script that interleaves LLM calls with control flow cannot be rewritten by an optimizer without understanding the programmer's intent; a declarative specification makes that intent explicit in the IR. The separation of specification (\texttt{.spl}) from execution (adapter/kernel) is what makes this possible: the runtime is free to choose \emph{how} to execute because the source only declares \emph{what} to execute.

The same \texttt{.spl} source compiles to idiomatic code in four target frameworks via \texttt{spl3\ splc} (\texttt{-\/-target\ langgraph}, \texttt{go}, \texttt{typescript}, \texttt{pocketflow}). The compiler reads the AST directly and maps each construct to the target's natural idiom --- \texttt{StateGraph} nodes for LangGraph, goroutines for Go parallel branches, \texttt{Promise.all} for TypeScript concurrency. This cross-framework compilation demonstrates that the declarative AST is a genuine intermediate representation, not a Python-specific wrapper: the same workflow specification produces native asynchronous code in each target's natural concurrency model. Full compilation examples and runtime details (kernel substrates, adapter bootstrap, template resolution, ecosystem tooling) are provided in Appendices B, F, and G.

\section{Empirical Evaluation}\label{empirical-evaluation}

\subsection{Experiment Design}\label{experiment-design}

We evaluate SPL's deterministic integration using a controlled dual-arm experiment on symbolic mathematics. The experiment covers verifier rungs R1 (SymPy) and R2 (SageMath); R3 (Lean 4) is demonstrated as a design contribution with passing unit tests but is not included in this grid (see C3 in \S\ref{our-contributions}). The benchmark consists of 20 problems spanning six difficulty tiers (T0--T5), split across two symbolic backends:

\begin{center}
\begin{tabular}{@{}p{0.6cm}p{1.3cm}p{5.5cm}p{6.0cm}@{}}
\toprule
Tier & Backend & Category & Example \\
\midrule
T0 & SymPy & Single-step polynomial & $d/dx\,(x^4 - 2x^2 + 1)$ \\
T1 & SymPy & Multi-step polynomial chains & expand $\to$ factor $\to$ solve \\
T2 & SymPy & Transcendental / limits / series / trig & $\sin(x)/x$ as $x\to 0$; Taylor $\sin(x)$ deg 5 \\
T3 & Sage  & Integration / linear systems / eigenvalues & $\int\!\sqrt{4-x^2}\,dx$; eigenvalues of $[[1,2],[3,4]]$ \\
T4 & Sage  & Laplace transforms / ODEs / summation / roots & $\mathcal{L}\{e^{-2t}\}$; $y'=y,\,y(0)=1$; $\sum 1/n^2$ \\
T5 & Sage  & Expert ODE + transform verification & $y''-3y'+2y=0$;\; $\mathcal{L}^{-1}\{s/(s^2+4)\}$ + verify \\
\bottomrule
\end{tabular}
\end{center}

T0--T2 problems (10 total) route to the SymPy kernel; T3--T5 (10 total) route to the SageMath kernel. The backend is a per-problem parameter; the \texttt{.spl} workflow is identical across both.

Ten models were evaluated: \texttt{sonnet-4-6} (Anthropic, cloud API), \texttt{gemma3} and \texttt{gemma4:e2b} (Google, local via Ollama), \texttt{qwen2.5} (Alibaba, local), \texttt{deepseek-v2:16b} (DeepSeek, local), \texttt{phi3} and \texttt{phi4} (Microsoft, local), \texttt{llama3.2} (Meta, local), \texttt{lfm2.5} (Liquid AI, local), and \texttt{rnj-1} (an experimental local model included to test the lower end of the capability spectrum). Models with mandatory extended chain-of-thought (qwen3, deepseek-r1) were excluded: the thinking trace exhausts the token budget before any structured output is emitted, making them fundamentally incompatible with the \texttt{expr\textbar{}op} contract (see Appendix H). Each of the ten models ran all 20 problems under two arms:

\begin{itemize}
\tightlist
\item
  \textbf{Solver arm} (\texttt{enable\_solver=true}): the model decomposes the problem into \texttt{expr\textbar{}op} steps; the symbolic kernel (SymPy or Sage) executes each step exactly; the verified chain is fed back to the model for narration. Pass = status \texttt{complete} (all steps kernel-verified end-to-end).
\item
  \textbf{LLM-only arm} (\texttt{enable\_solver=false}): the model solves and narrates directly with no kernel involvement. Pass = non-empty response returned (status \texttt{complete} or \texttt{unverified\_success}). This arm measures \emph{output production} --- the model's ability to generate a complete response --- not mathematical correctness. Mathematical accuracy is deliberately left unverified: the comparison isolates what SPL's deterministic mode adds (machine-verified correctness) against a baseline that represents what the LLM produces on its own.
\end{itemize}

Two sessions were run: a pilot (\texttt{exp-20260615-073849}, r=1, 400 cells) and a repeated run (\texttt{exp-20260615-191224}, r=3, 1200 cells). Results below are from the r=3 session; the pilot is retained in Appendix E for comparison.

\subsection{Results}\label{results}

\textbf{Pass rate and latency by model and arm} (mean over 3 runs/cell, sorted by solver pass rate; 95\% bootstrap CI on solver pass rate, 10k resamples):

\begin{center}
\begin{tabular}{@{}lrrrlrrr@{}}
\toprule
 & \multicolumn{4}{c}{Pass rate (\%)} & \multicolumn{3}{c}{Latency} \\
\cmidrule(lr){2-5} \cmidrule(l){6-8}
Model & LLM-only & Solver & 95\% CI & $\Delta$ & LLM-only & Solver & $\Delta$ \\
\midrule
gemma4:e2b      & 97  & 93 & [87, 98] & $-3$  & 11.9s & 14.8s & $+25\%$ \\
sonnet-4-6      & 100 & 85 & [75, 93] & $-15$ & 13.1s &  9.7s & $-25\%$ \\
rnj-1           & 100 & 82 & [72, 90] & $-18$ &  7.4s &  6.4s & $-13\%$ \\
gemma3          & 100 & 73 & [62, 83] & $-27$ &  4.7s &  4.6s & $ -1\%$ \\
qwen2.5         & 100 & 72 & [60, 83] & $-28$ &  8.6s &  3.9s & $-55\%$ \\
phi4            & 100 & 67 & [55, 78] & $-33$ & 27.3s & 12.1s & $-56\%$ \\
llama3.2        & 100 & 65 & [53, 77] & $-35$ &  5.0s &  3.0s & $-41\%$ \\
deepseek-v2:16b & 100 & 53 & [40, 65] & $-47$ & 16.8s &  7.9s & $-53\%$ \\
lfm2.5          &  77 & 37 & [25, 50] & $-40$ &  7.1s & 12.2s & $+71\%$ \\
phi3            & 100 & 32 & [20, 43] & $-68$ &  7.4s &  4.1s & $-45\%$ \\
\bottomrule
\end{tabular}
\end{center}

\begin{figure}
\centering
\includegraphics{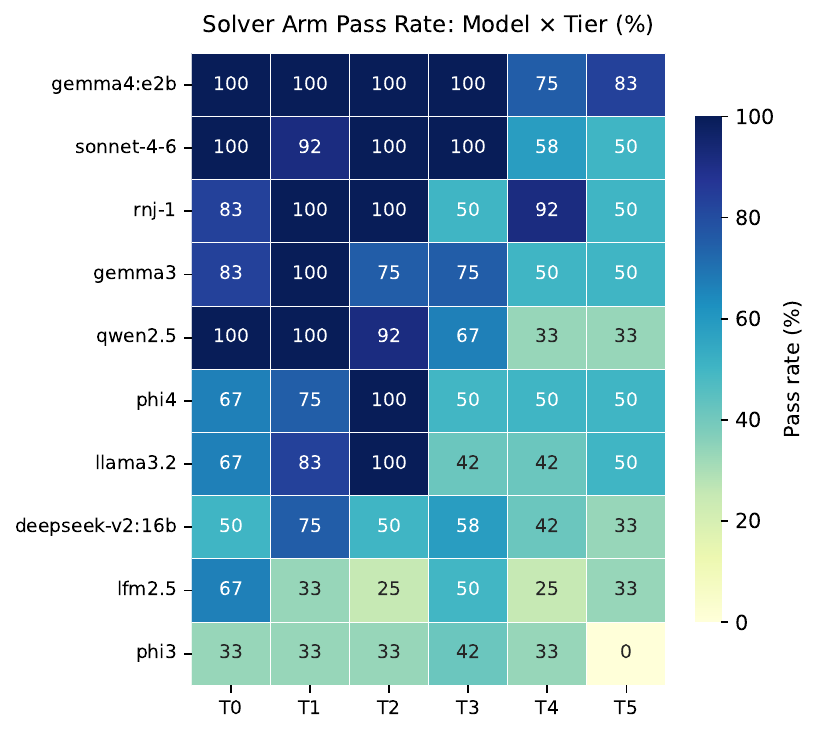}
\caption{Solver Arm Pass Rate: Models (rows, sorted by aggregate) $\times$ tiers (columns). Diagonal gradient shows capability hierarchy; vertical gradient shows backend difficulty (SymPy 78.5\% vs Sage 54.9\%).}
\end{figure}

LLM-only ``pass'' = non-empty response (unverified). Solver ``pass'' = complete symbolic chain (machine-verified by SymPy or Sage kernel). The two arms measure different things: output production vs.~verified correctness.

\subsection{Findings}\label{findings}

\textbf{F1 --- Output-fence cleaning eliminated plan\_format\_error entirely.} In a preliminary run, phi4 produced its decomposition plans wrapped in markdown fences (\texttt{\textasciigrave{}\textasciigrave{}\textasciigrave{}plaintext\ ...\ \textasciigrave{}\textasciigrave{}\textasciigrave{}}), causing SyntaxErrors in the kernel and 0\% solver pass. Adding \texttt{strip\_fences()} --- a regex-based fence stripper applied immediately after \texttt{decompose\_problem()} --- resolved this. Across both the 400-cell pilot and the 1200-cell repeated run, zero failures were attributed to \texttt{plan\_format\_error} in any model. The dominant failure mode is \texttt{solver\_error} (kernel-rejected expressions): \emph{surface} format compliance (markdown fences, whitespace) is not the bottleneck --- \emph{semantic} expression correctness at the kernel boundary is. Models that fail are not failing to produce \texttt{expr\textbar{}op}-shaped text; they are producing expressions the symbolic engine cannot evaluate (wrong variable names, unsupported operations, malformed syntax).

\textbf{F2 --- solver\_error is the primary failure mode; lfm2.5 is the plan\_error outlier.} Every failure in the solver arm is \texttt{solver\_error} or \texttt{plan\_error} --- never \texttt{plan\_format\_error}. Models that fail are generating syntactically valid \texttt{expr\textbar{}op} plans but producing expressions the kernel cannot evaluate: wrong variable names, unsupported operations for the active backend, or malformed expression syntax (e.g., \texttt{\^{}} instead of \texttt{**}). \texttt{lfm2.5} is the outlier: 26/60 solver failures are \texttt{plan\_error} (invalid step structure), not \texttt{solver\_error}, indicating a different failure mode --- plan generation rather than expression evaluation.

\textbf{F3 --- gemma4:e2b achieves near-solver parity; sonnet-4-6 is the most stable model.} \texttt{gemma4:e2b} scores 93\% solver vs 97\% LLM-only --- the closest any model comes to zero-cost verified correctness. \texttt{sonnet-4-6} scores exactly 85\% in both the r=1 pilot and the r=3 repeated run, showing stable results. \texttt{rnj-1} drops from 90\% (pilot) to 82\% (repeated), revealing pilot variance. \texttt{phi4} shows the largest pilot inflation: 85\% (r=1) → 67\% (r=3), an 18-point drop indicating the value of repeated runs for ranking stability.

\textbf{F4 --- Sage problems are structurally harder for the solver arm, with a confound.} SymPy problems (T0--T2): 78\% aggregate solver pass rate (95\% CI: {[}73, 82{]}). Sage problems (T3--T5): 54\% aggregate solver pass rate (95\% CI: {[}48, 60{]}). The Sage backend exposes harder decomposition demands: Laplace transforms, first- and second-order ODEs, infinite sums, and inverse transform verification require the model to correctly name Sage-specific operations (\texttt{laplace}, \texttt{inverse\_laplace}, \texttt{solve\_system}) and format expressions the Sage kernel can parse. T4 (50\%) and T5 (43\%) are the hardest tiers across all models. \textbf{Confound:} the SymPy/Sage gap conflates two factors --- Sage is a less familiar API surface for most models \emph{and} is assigned the intrinsically harder mathematics (T3--T5). Disentangling the two would require running overlapping problems through both backends. Per-tier numbers (T0--T5) should be read as illustrative of the difficulty gradient rather than statistically precise estimates --- each tier contains 2--4 problems, yielding 6--12 runs per model-tier cell. The paper's primary contribution is the two-mode language design; the experiment demonstrates the architecture's feasibility and surfaces the format-compliance hierarchy, not an exhaustive benchmark of model capabilities.

\textbf{F5 --- Solver arm accelerates most models, but early-exit confounds the comparison.} Seven models are faster in the solver arm, led by \texttt{qwen2.5} (-55\%), \texttt{phi4} (-56\%), \texttt{deepseek-v2:16b} (-53\%). The kernel short-circuits LLM chain-of-thought: once decomposition is done, the symbolic engine resolves the chain without further LLM calls, and early-exit on \texttt{solver\_error} skips the narration call entirely. However, this means part of the speedup is an artifact of \emph{failing faster} --- a \texttt{solver\_error} exits before the narration \texttt{GENERATE}, producing lower mean latency for lower-scoring models. A cleaner comparison would report latency on \emph{passing runs only}; we report all-run means here for completeness. Additionally, \texttt{sonnet-4-6} latency is measured through \texttt{claude\_cli} (shell invocation) while local models use direct HTTP, so cross-model latency comparisons are not instrument-normalized. \texttt{gemma4:e2b} (+25\%) and \texttt{lfm2.5} (+71\%) are slower --- \texttt{gemma4:e2b} because high accuracy means it completes chains and pays the full narration call; \texttt{lfm2.5} because its 26 plan\_errors require longer retry attempts before failing.

The SPL status codes (\texttt{plan\_error}, \texttt{solver\_error}, \texttt{complete}) make per-category failure analysis automatic --- each cell records its failure mode, enabling per-tier, per-backend, and per-model breakdown without log parsing.

\subsection{Discussion of Evaluation Design}\label{discussion-of-evaluation-design}

\textbf{Pass/fail criterion and baseline asymmetry.} The solver arm pass criterion is machine-verified: status \texttt{complete} means the symbolic kernel executed every decomposed step without error, producing a chain-verified result. The LLM-only arm pass criterion is deliberately weaker: status \texttt{complete} or \texttt{unverified\_success} means the model produced a non-empty response. This asymmetry is by design --- the experiment does not claim ``the solver arm is more accurate than the LLM-only arm.'' Instead, it measures the \emph{cost of verified correctness}: given that the LLM can produce output (near-100\% in the LLM-only arm), how much does adding machine-verification via the solver reduce the pass rate? The answer --- from 3\% (gemma4:e2b) to 68\% (phi3) --- reveals the format-compliance hierarchy.

\textbf{Why scoring LLM-only accuracy is itself hard.} A natural follow-up question is: how mathematically correct are the LLM-only outputs? We attempted post-hoc verification by extracting answers from LLM narrations and comparing against the solver arm's verified results using SymPy \texttt{simplify}. The attempt yielded unreliable scores (3--28\% match rate) --- not because the models are necessarily wrong, but because LLM outputs are \emph{representationally polymorphic}: the same correct derivative may appear as \texttt{4x³\ -\ 4x} (Unicode), \texttt{4*x**3\ -\ 4*x} (Python), \texttt{\$4x\^{}3-4x\$} (LaTeX), or ``four x cubed minus four x'' (prose). No heuristic extraction reliably recovers a canonical symbolic form from free-text narration, and even human reviewers struggle to verify correctness across these representations without a computational tool.

This difficulty is not a limitation of the experiment --- it is the central observation the paper makes. The solver arm eliminates representational ambiguity entirely: the kernel operates on canonical symbolic expressions and returns machine-checkable results. The fact that \emph{scoring} LLM-only mathematical accuracy requires either a human expert or a symbolic verifier is itself evidence for the two-mode architecture SPL proposes.

\textbf{Repeated runs and ranking stability.} The r=1 pilot (400 cells) and r=3 session (1200 cells) tell a consistent story at the top and bottom of the ranking but diverge in the middle. The top-3 ranking (gemma4:e2b → sonnet-4-6 → rnj-1) and bottom-2 (lfm2.5, phi3) are stable across both sessions. The middle tier shows pilot variance of up to 18 percentage points (phi4: 85\% → 67\%), indicating that r=3 is necessary for reliable ranking. Full per-run comparison is in Appendix E.5.

\textbf{Thinking-mode exclusion and CoT compatibility.} Two model families (\texttt{qwen3}, \texttt{deepseek-r1}) were excluded pre-experiment: they run mandatory extended chain-of-thought that exhausts the token budget before emitting structured output, violating the \texttt{expr\textbar{}op} contract. The experiment uses non-thinking alternatives (\texttt{deepseek-v2:16b}, \texttt{llama3.2}) instead.

This exclusion reflects an architectural insight, not merely a token-budget constraint. SPL intentionally offloads mathematical reasoning to the deterministic kernel (System~2), reducing the LLM's role (System~1) to that of a \emph{format translator} --- mapping natural-language problems into strict \texttt{expr\textbar{}op} syntax. Extended-thinking models are optimized for open-ended, verbose deliberation, which conflicts with the concise syntactic formatting SPL requires. The empirical results confirm this design: \texttt{gemma4:e2b}, a \textasciitilde2B-parameter open-source model, achieves 93\% verified correctness --- not because it reasons mathematically (the kernel does that), but because it reliably produces the structured format the kernel requires. For SPL's two-mode architecture, a fast, format-compliant model is the optimal tool; extended reasoning is redundant work the kernel will perform deterministically.

SPL is not inherently incompatible with chain-of-thought reasoning --- a thinking model's hidden reasoning trace could be captured as a distinct \texttt{GENERATE} step whose output is discarded before the structured \texttt{expr\textbar{}op} extraction step. Models with \emph{optional} thinking modes (e.g., \texttt{qwen3} with \texttt{-\/-no-think}) would work with SPL as-is by disabling the thinking trace at invocation time.

\section{Discussion}\label{discussion}

\subsection{Completing the Computation Matrix}\label{the-22-matrix-closed}

Table 2 situates SPL against existing systems on the computation matrix from Section~\ref{introduction}:

\begin{longtable}[]{@{}
  >{\raggedright\arraybackslash}p{(\columnwidth - 8\tabcolsep) * \real{0.2000}}
  >{\raggedright\arraybackslash}p{(\columnwidth - 8\tabcolsep) * \real{0.2000}}
  >{\raggedright\arraybackslash}p{(\columnwidth - 8\tabcolsep) * \real{0.2000}}
  >{\raggedright\arraybackslash}p{(\columnwidth - 8\tabcolsep) * \real{0.2000}}
  >{\raggedright\arraybackslash}p{(\columnwidth - 8\tabcolsep) * \real{0.2000}}@{}}
\toprule\noalign{}
\begin{minipage}[b]{\linewidth}\raggedright
System
\end{minipage} & \begin{minipage}[b]{\linewidth}\raggedright
Probabilistic (design)
\end{minipage} & \begin{minipage}[b]{\linewidth}\raggedright
Probabilistic (runtime)
\end{minipage} & \begin{minipage}[b]{\linewidth}\raggedright
Deterministic (design)
\end{minipage} & \begin{minipage}[b]{\linewidth}\raggedright
Deterministic (runtime)
\end{minipage} \\
\midrule\noalign{}
\endhead
\bottomrule\noalign{}
\endlastfoot
LangGraph & \checkmark & \checkmark & --- & --- \\
AutoGen & \checkmark & \checkmark & --- & --- \\
DSPy & \checkmark & \checkmark & --- & --- \\
AlphaGeometry & --- & \checkmark & \checkmark & \checkmark (geometry only) \\
\textbf{SPL} & \textbf{\checkmark} & \textbf{\checkmark} & \textbf{\checkmark} & \textbf{\checkmark (any domain)} \\
\end{longtable}

No prior workflow language fills all four cells for arbitrary domains. AlphaGeometry fills all four but is domain-locked to geometry. SPL fills all four for any domain with a Python-callable verifier, with the boundary between cells declared in source rather than implicit in framework internals. The programmer writes one \texttt{.spl} file; the runtime resolves which cells are active at invocation --- a difference of kind, not degree. Correctness-vs-speed trade-offs become deployment decisions, not source-embedded choices.

\subsection{The System 1/2 Decomposition}\label{the-system-12-decomposition}

SPL's two-mode architecture maps onto dual-process theory: \emph{probabilistic/intuitive} (System 1, LLM) versus \emph{deterministic/mechanical} (System 2, kernel). System 1 reads problems, chooses approaches, and narrates results; System 2 executes algebra, checks identities, and certifies proofs. The empirical results (Finding F1) expose a System 1/2 boundary mismatch: models with strong System 1 reasoning (100\% LLM-only pass) fail at the System 1 format-mapping task SPL requires --- translating problem statements into structured \texttt{expr\textbar{}op} plans. This surfaces a hierarchy: reasoning ability → format compliance → verifier access.

\subsection{Privacy and Edge Deployment}\label{privacy-and-edge-deployment}

A significant advantage of SPL's DODA principle is its enablement of privacy-preserving, edge-deployed workflows. In educational contexts --- such as the student homework assistant motivating our design --- data privacy regulations (FERPA, GDPR) are paramount. Because the \texttt{.spl} specification is entirely agnostic to the execution backend, workflows can be deployed to local environments using adapters like Ollama. This allows institutions to run fully offline, zero-marginal-cost neurosymbolic pipelines where student data never leaves the local device, while still benefiting from machine-verified correctness via the local IPython kernel.

The 1200-run experiment demonstrates this concretely: nine of the ten evaluated models ran locally via Ollama with zero API token cost, and the local SymPy/SageMath kernels required no network access. The solver arm's multi-call pattern (decompose, kernel execution, narrate) adds \textasciitilde2--3 LLM calls relative to a single-shot LLM response, but for local models this cost is measured in seconds of local compute rather than API dollars. While local deployment shifts the cost from API tokens to hardware requirements (e.g., GPU VRAM for larger models), the trade-off is highly favorable for privacy-sensitive applications where verified correctness and data sovereignty are non-negotiable.

\subsection{The Structured Output Bottleneck}\label{the-structured-output-bottleneck}

The format-compliance gap (Finding F1) extends beyond symbolic mathematics: any structured-output contract exposes a capability hierarchy where models lose verifier access if they cannot emit the required format, regardless of reasoning ability. A future feature could probe format compliance at runtime and route incapable models to the LLM-only arm; \texttt{EVALUATE} already provides the dispatch mechanism.

\subsection{Pluggability vs.~AlphaGeometry-Style Integration}\label{pluggability-vs.-alphageometry-style-integration}

AlphaGeometry {[}16{]} achieves state-of-the-art geometry theorem proving by hard-coding the neuro-symbolic loop: the neural model generates proof candidates, the symbolic engine (DD-Geometry) verifies them, and the system iterates. This tight coupling is appropriate for a single-domain system optimized for competition performance.

SPL takes the opposite design stance: the symbolic component is always an external, pluggable tool registered via \texttt{CREATE\ TOOL\_API} or \texttt{@spl\_tool}. This generality trades peak single-domain performance for domain-independence. A domain expert adds a new verifier in Python; the SPL workflow language and runtime require no modification. The same \texttt{ASSERT} semantics that check a SymPy polynomial identity can check a NetworkX graph property, a Z3 satisfiability result, or a scipy hypothesis test --- any predicate the domain expert can express in Python. Empirically, open-source models like gemma4:e2b achieve 93\% verified correctness on symbolic mathematics --- not because the model reasons mathematically (the kernel does that), but because it reliably produces the structured \texttt{expr\textbar{}op} format the deterministic engine requires. This is precisely the design's thesis: mathematical reasoning is offloaded to the kernel, making \emph{format-mapping capability} --- not model scale or proprietary access --- the bottleneck. When a \textasciitilde2B-parameter open-source model outperforms a frontier model on verified correctness, it demonstrates that the two-mode architecture makes high-quality verified computation accessible to everyone.

\subsection{Limitations}\label{limitations}

\textbf{Kernel startup overhead.} The IPython kernel adds \textasciitilde2 seconds of cold-start latency per run. For short workflows with a single \texttt{SOLVE} step, this overhead is disproportionate. The \texttt{KernelSession} substrate (in-process \texttt{exec()}) is a lower-overhead alternative but lacks the full scientific Python environment of the Jupyter kernel.

\textbf{Verifier ladder setup cost.} R2 (SageMath) and R3 (Lean 4) require non-trivial installation beyond \texttt{pip\ install\ spl-llm}. SageMath requires a separate conda package or system install; Lean 4 requires the Elan toolchain and Mathlib download (\textasciitilde2 GB). These are not barriers for a practitioner setting up a proof-carrying pipeline, but they mean the full verifier ladder is not available in a zero-configuration deployment.

\textbf{Pass oracle.} The r=3 results in this paper use SPL status codes as the pass oracle: a solver-arm cell passes when the symbolic kernel executes all decomposed steps without error (status \texttt{complete}). The LLM sanity gate (an independent \texttt{sonnet-4-6} judge, described in Appendix E.6) was used as an additional validation check in the 400-run pilot, not as the primary oracle. Replacing the judge with a deterministic SymPy ground-truth check of the LLM-only arm's mathematical accuracy would give a cleaner correctness comparison and is planned for the next experiment round.

\subsection{Future Work}\label{future-work}

\textbf{Format-compliance routing.} A pre-flight probe checking whether the active model can emit the required format would gate the solver arm and eliminate format-error failures.

\textbf{Query optimizer.} SPL's declarative semantics enable a runtime optimizer: merging redundant \texttt{GENERATE} calls, caching \texttt{SELECT} results, model downgrading for cheap sub-tasks --- all without modifying the \texttt{.spl} source.

\textbf{Proof-carrying workflows.} R3 (Lean 4) currently generates a proof badge but does not embed the proof artifact in the workflow's output. A proof-carrying extension would attach the Lean 4 certificate as a typed field in the \texttt{RETURN} payload, enabling downstream consumers to independently re-verify without re-running the workflow.

\textbf{Distributed verifier routing.} The Momagrid adapter {[}22{]} currently routes LLM calls to worker nodes. The same routing mechanism could route \texttt{SOLVE} dispatches to verifier-specialized nodes (a node with a GPU-accelerated SageMath install, or a node with Lean 4 and Mathlib pre-warmed), enabling a distributed verifier ladder over the existing Momagrid infrastructure.

\section{Conclusion}\label{conclusion}

Most orchestration frameworks today excel at the probabilistic side --- LLM invocation, multi-agent routing, chain-of-thought scaffolding --- with layers of abstraction built over decades of LLM research. The deterministic side, however, remains fragmented: symbolic computation (algebra, geometry, formal proof) lives in separate tools (SymPy, SageMath, Lean) with no native integration path. When practitioners need both --- a workflow that decomposes with an LLM, verifies with a solver, and proves with Lean --- they resort to custom Python glue that is neither declarative nor portable.

SPL closes this gap through a unified declarative approach grounded in an epistemic --- not temporal --- reading of dual-process framing: System 1 (probabilistic, intuitive) and System 2 (deterministic, mechanical) are distinguished by their mode of computation, not their speed. A single \texttt{.spl} specification expresses both modes in the same syntax, with the same variable bindings. The runtime handles mode routing; the programmer declares \emph{what} each step computes, not \emph{how} or \emph{where}. This separation enables the DODA principle: a workflow authored once runs unchanged across Ollama (local), OpenRouter (cloud), or Momagrid (distributed), with model and verifier selection deferred to invocation time.

We validate SPL through two complementary lenses. An extensive 78-recipe cookbook illustrates expressiveness across 12 workflow categories, from basic Q\&A to formal proof verification. A 1200-run controlled experiment ($10 \times 20 \times 2 \times 3 = 1200$ runs; with problems spanning 6 difficulty tiers) on symbolic mathematics reveals a structural finding: the most capable models (gemma4:e2b: 93\%, sonnet-4-6: 85\%) achieve near-solver parity with the LLM-only output-production baseline, demonstrating that verified computation carries near-zero runtime cost. Format compliance --- the ability to emit structured \texttt{expr\textbar{}op} plans --- emerges as a separable skill from mathematical reasoning ability; models that fail format translation lose verifier access entirely, regardless of reasoning strength. The verifier ladder (SymPy for basic algebra, SageMath for advanced mathematics, Lean 4 for formal proofs) makes correctness guarantees a runtime parameter, not a specification artifact.

A broader implication emerges: declarative composition fundamentally shifts the LLM's role from mathematical reasoner to format translator. The kernel handles derivation; the LLM handles decomposition and narration. This lowers the capability bar --- a small, fast, open-source model suffices when format compliance, not reasoning depth, is the bottleneck --- and enables fully offline, privacy-preserving deployment via local adapters like Ollama, where student data never leaves the device yet still benefits from machine-verified correctness.

These findings extend beyond symbolic mathematics. A domain requiring both generative reasoning and deterministic verification --- code generation and testing, scientific computation with certification, knowledge graphs with logical constraints, JSON schema validation --- can benefit from our declarative two-mode composition. SPL is released under Apache 2.0. Source code, experimental data, and reproducibility artifacts are all publicly available at \url{https://github.com/digital-duck/SPL.py}.

\section*{References}\label{references}

{[}1{]} Q. Wu et al., ``AutoGen: Enabling Next-Gen LLM Applications via Multi-Agent Conversation,'' arXiv:2308.08155, August 2023.

{[}2{]} J. Moura, ``CrewAI: Framework for Orchestrating Role-Playing Autonomous AI Agents,'' 2024.

{[}3{]} M. Stonebraker and G. Kemnitz, ``The POSTGRES Next Generation Database Management System,'' CACM, 1991.

{[}4{]} Oracle Corporation, ``PL/SQL User's Guide and Reference,'' 1992.

{[}5{]} L. Beurer-Kellner et al., ``Prompting Is Programming: A Query Language for Large Language Models,'' arXiv:2212.06094, PLDI 2023.

{[}6{]} O. Khattab et al., ``DSPy: Compiling Declarative Language Model Calls into Self-Improving Pipelines,'' arXiv:2310.03714, ICLR 2024.

{[}7{]} L. Zheng et al., ``SGLang: Efficient Execution of Structured Language Model Programs,'' arXiv:2312.07104, NeurIPS 2024.

{[}8{]} S. Amini, Y. Benajiba et al., ``Open Agent Specification (Agent Spec): A Unified Representation for AI Agents,'' arXiv:2510.04173, October 2025.

{[}9{]} I. Daunis, ``A Declarative Language for Building and Orchestrating LLM-Powered Agent Workflows,'' arXiv:2512.19769, December 2025.

{[}10{]} A. M. Gliozzo, J. Lee, and N. Defosse, ``Agentics 2.0: Logical Transduction Algebra for Agentic Data Workflows,'' arXiv:2603.04241, March 2026.

{[}11{]} G. Trooskens, A. Karlsberg, A. Sharma et al., ``Compiled AI: Deterministic Code Generation for LLM-Based Workflow Automation,'' arXiv:2604.05150, April 2026.

{[}12{]} L. Qiu, Y. Ye, and Z. Gao, ``Blueprint First, Model Second: A Framework for Deterministic LLM Workflow,'' arXiv:2508.02721, August 2025.

{[}13{]} B. Liu et al., ``LLM+P: Empowering Large Language Models with Optimal Planning Proficiency,'' arXiv:2304.11477, April 2023.

{[}14{]} S. Szeider, ``MCP-Solver: Integrating Language Models with Constraint Programming Systems,'' arXiv:2501.00539, SAT 2025.

{[}15{]} K. Hua, D. Wang, Y. Gu, and X. Ma, ``DUPLEX: Agentic Dual-System Planning via LLM-Driven Information Extraction,'' arXiv:2603.23909, March 2026.

{[}16{]} T. Trinh et al., ``Solving Olympiad Geometry Without Human Demonstrations,'' Nature 625(7995):476--482, 2024.

{[}17{]} K. Yang et al., ``LeanDojo: Theorem Proving with Retrieval-Augmented Language Models,'' arXiv:2306.15626, NeurIPS 2023.

{[}18{]} G. Lample et al., ``HyperTree Proof Search for Neural Theorem Proving,'' arXiv:2205.11491, NeurIPS 2022.

{[}19{]} M. Chen et al., ``Evaluating Large Language Models Trained on Code,'' arXiv:2107.03374, July 2021.

{[}20{]} Y. Li et al., ``Competition-Level Code Generation with AlphaCode,'' arXiv:2203.07814, Science 2022.

{[}21{]} D. Kahneman, ``Thinking, Fast and Slow,'' Farrar, Straus and Giroux, 2011.

{[}22{]} W. G. Gong, ``Momagrid: A Decentralized Inference Runtime with Linear Complexity via Semantic Chunking,'' unpublished manuscript, 2026.

{[}23{]} ACM, ``Compositional AI Beyond LLMs: System Implications of Neuro-Symbolic-Probabilistic Architectures,'' in Proceedings of the 52nd Annual ACM SIGPLAN Symposium on Principles and Practice of Programming Languages (ASPLOS 2026), ACM DL: 10.1145/3760250.3762235, April 2026.

{[}24{]} N. Schuler, V. Scotti, and R. Mirandola, ``Beyond Monolithic Models: Symbolic Seams for Composable Neuro-Symbolic Architectures,'' arXiv:2603.15087, March 2026.

\clearpage

\section*{Appendix A: Extended Backus-Naur Form (EBNF) Grammar}\label{appendix-a-extended-backus-naur-form-ebnf-grammar}

The grammar below covers the SPL language. Terminal symbols are quoted strings; non-terminals are lowercase identifiers. \texttt{*} = zero or more, \texttt{+} = one or more, \texttt{?} = optional, \texttt{\textbar{}} = alternation. The \texttt{\{@var\}} interpolation syntax (used inside string expressions for kernel dispatch) is handled at the executor level and is not shown separately.

\begin{verbatim}
(* Top-level structure *)
program         ::= (import_stmt | workflow_def | procedure_def
                     | create_fn_stmt | create_tool_stmt)*

import_stmt     ::= 'IMPORT' string_literal

workflow_def    ::= 'WORKFLOW' identifier
                    input_decl? output_decl? 'DO'
                    stmt_list
                    'END'

procedure_def   ::= 'PROCEDURE' identifier
                    input_decl? output_decl? 'DO'
                    stmt_list
                    'END'

input_decl      ::= 'INPUT' ':'? param_spec (',' param_spec)*
output_decl     ::= 'OUTPUT' ':'? param_spec (',' param_spec)*
param_spec      ::= '@'identifier type_ann? ('DEFAULT' expr)?
type_ann        ::= 'TEXT' | 'LIST' | 'SET' | 'INT' | 'FLOAT' | 'BOOL'
                  | 'IMAGE' | 'AUDIO' | 'VIDEO'

(* Statements *)
stmt_list       ::= stmt*
stmt            ::= generate_stmt
                  | evaluate_stmt
                  | while_stmt
                  | exception_stmt
                  | solve_stmt
                  | assert_stmt
                  | call_stmt
                  | call_parallel_stmt
                  | return_stmt
                  | logging_stmt
                  | assign_stmt

(* Probabilistic primitives *)
generate_stmt   ::= 'GENERATE' call_expr 'INTO' '@'identifier
                    select_clause? prompt_clause? budget_clause?

budget_clause   ::= 'WITH' 'OUTPUT' 'BUDGET' integer 'TOKENS'

select_clause   ::= 'SELECT' select_item (',' select_item)*
select_item     ::= system_role_call
                  | '@'identifier 'AS' identifier
                  | string_literal 'AS' identifier
                  | rag_call
                  | memory_call

prompt_clause   ::= 'PROMPT' string_literal

system_role_call ::= 'system_role' '(' string_literal ')'
rag_call        ::= 'rag' '(' string_literal (',' expr)? ')'
memory_call     ::= 'memory' '(' string_literal ')'

evaluate_stmt   ::= 'EVALUATE' '@'identifier
                    when_clause+ else_clause? 'END'
when_clause     ::= 'WHEN' condition 'THEN' stmt_list
else_clause     ::= 'ELSE' stmt_list
condition       ::= '=' string_literal          (* deterministic equality *)
                  | string_literal              (* semantic / LLM-judge *)
                  | 'contains' '(' string_literal ')'
                  | 'startswith' '(' string_literal ')'
                  | python_expr                 (* deterministic Python bool *)

while_stmt      ::= 'WHILE' condition 'DO' stmt_list 'END'

exception_stmt  ::= 'EXCEPTION' exception_clause+ 'END'
exception_clause ::= 'WHEN' exception_type 'THEN' stmt_list
exception_type  ::= 'HallucinationDetected' | 'ContextLengthExceeded'
                  | 'BudgetExceeded' | 'ModelRefused' | 'TimeoutError'
                  | 'WorkflowCompositionError' | identifier

(* Deterministic primitives *)
solve_stmt      ::= 'SOLVE' '@'identifier type_ann? kernel_hint? ':=' python_expr
kernel_hint     ::= 'SYMPY' | 'SAGE' | 'LEAN' | 'PYTHON'

assert_stmt     ::= 'ASSERT' python_expr
                    ('OTHERWISE' stmt_list)?

(* Workflow composition *)
call_stmt       ::= 'CALL' call_expr 'INTO' '@'identifier
call_expr       ::= identifier '(' arg_list? ')'
arg_list        ::= arg (',' arg)*
arg             ::= '@'identifier | string_literal | named_arg
named_arg       ::= identifier '=' expr

call_parallel_stmt ::= 'CALL' 'PARALLEL' call_stmt+ 'END'

(* Output and side effects *)
return_stmt     ::= 'RETURN' expr
                    ('WITH' identifier '=' expr (',' identifier '=' expr)*)?
                    (* COMMIT is a deprecated alias for RETURN *)

logging_stmt    ::= 'LOGGING' expr ('LEVEL' log_level)?
log_level       ::= 'INFO' | 'DEBUG' | 'WARNING' | 'ERROR'

assign_stmt     ::= '@'identifier ':=' expr

(* Tool and function registration *)
create_fn_stmt  ::= 'CREATE' 'FUNCTION' identifier
                    '(' param_list? ')'
                    ('RETURNS' | 'RETURN') type_ann
                    'AS' '$$' string_literal '$$'

create_tool_stmt ::= 'CREATE' 'TOOL_API' identifier
                     '(' param_list? ')'
                     ('RETURNS' | 'RETURN') type_ann
                     'AS' 'PYTHON' '$$' string_literal '$$'

param_list      ::= param_spec (',' param_spec)*

(* Expressions *)
expr            ::= '@'identifier
                  | string_literal
                  | fstring_literal
                  | number_literal
                  | bool_literal
                  | none_literal
                  | list_literal
                  | map_literal
                  | binary_expr
                  | unary_expr
                  | call_expr

fstring_literal ::= 'f"' (char | '{' '@'identifier '}')* '"'
list_literal    ::= '[' (expr (',' expr)*)? ']'
map_literal     ::= '{' (string_literal ':' expr (',' string_literal ':' expr)*)? '}'
binary_expr     ::= expr ('+'|'-'|'*'|'/'|'=='|'!='|'<'|'>'|'<='|'>='
                          |'AND'|'OR') expr
unary_expr      ::= 'NOT' expr
python_expr     ::= string_literal              (* dispatched to kernel *)

(* Lexical *)
identifier      ::= [a-zA-Z_][a-zA-Z0-9_]*
string_literal  ::= '"' [^"\\]* '"' | "'" [^'\\]* "'"
                  | "r'" [^']* "'" | 'r"' [^"]* '"'
                  | "'''" char* "'''" | '"""' char* '"""'
number_literal  ::= [0-9]+ ('.' [0-9]+)?
bool_literal    ::= 'true' | 'false' | 'True' | 'False'
none_literal    ::= 'None' | 'null'
comment         ::= ('--' | '#') char* newline
                  | '/*' char* '*/'
\end{verbatim}

String literals use double quotes as preferred, single quotes are accepted, but can interfere with keyword highlighting in the VS Code syntax highlighter. The \texttt{\{@var\}} interpolation in string literals sent to \texttt{SOLVE}/\texttt{ASSERT} is resolved by the executor before kernel dispatch.

Comments are stripped before parsing. Three styles are supported: line comments with \texttt{-\/-} or \texttt{\#} (until end of line), and block comments with \texttt{/*\ ...\ */} (multi-line).

\section*{Appendix B: Adapter System}\label{appendix-b-adapter-system}

\subsection*{B.1 Mandatory Adapters}\label{b.1-mandatory-adapters}

Four adapters are always available after \texttt{pip\ install\ spl-llm}. All support \texttt{-\/-llm\ \textless{}adapter\textgreater{}:\textless{}model\_id\textgreater{}} format. The legacy \texttt{-\/-adapter} and \texttt{-\/-model} flags remain available for backward compatibility.

\begin{longtable}[]{@{}
  >{\raggedright\arraybackslash}p{5cm}
  >{\raggedright\arraybackslash}p{8.5cm}@{}}
\toprule\noalign{}
\begin{minipage}[b]{\linewidth}\raggedright
Adapter:Model spec
\end{minipage} & \begin{minipage}[b]{\linewidth}\raggedright
Protocol
\end{minipage} \\
\midrule\noalign{}
\endhead
\bottomrule\noalign{}
\endlastfoot
\texttt{ollama:gemma3} & HTTP POST to \texttt{localhost:\ 11434/v1/chat/completions} (OpenAI-compatible) \\
\texttt{claude\_cli:sonnet-4-6} & Shells out to \texttt{claude} CLI; captures stdout \\
\texttt{openrouter:qwen/qwen3-8b} & HTTPS to \texttt{api.openrouter.ai}; requires \texttt{OPENROUTER\_API\_KEY} \\
\texttt{momagrid:gemma3} & POST \texttt{/tasks} on \texttt{MOMAGRID\_HUB\_URL}; poll \texttt{/tasks/\{id\}} for results \\
\end{longtable}

We choose \texttt{gemma} and \texttt{qwen} as default models due to their continuous open-source supports.

\subsection*{B.2 dd-llm Bridge (Cloud Providers)}\label{b.2-dd-llm-bridge-cloud-providers}

When the \texttt{dd-llm} package is installed, ten additional cloud providers are available through a unified bridge adapter. The bridge normalizes provider-specific authentication and response shapes behind the same \texttt{BaseAdapter} interface:

\begin{longtable}[]{@{}
  >{\raggedright\arraybackslash}p{2.5cm}
  >{\raggedright\arraybackslash}p{6.0cm}@{}}
\toprule\noalign{}
Provider & Model prefix example \\
\midrule\noalign{}
\endhead
\bottomrule\noalign{}
\endlastfoot
Anthropic & \texttt{anthropic/claude-sonnet-4-6} \\
OpenAI & \texttt{openai/gpt-4o} \\
Google & \texttt{google/gemini-2.0-flash} \\
Mistral & \texttt{mistral/mistral-large} \\
Cohere & \texttt{cohere/command-r-plus} \\
Together AI & \texttt{together/\allowbreak meta-llama/\allowbreak Llama-3-70b} \\
Groq & \texttt{groq/llama3-70b-8192} \\
Fireworks & \texttt{fireworks/\allowbreak accounts/\allowbreak fireworks/\allowbreak models/\allowbreak mixtral-8x22b} \\
Perplexity & \texttt{perplexity/sonar-large-32k-online} \\
DeepSeek & \texttt{deepseek/deepseek-chat} \\
\end{longtable}

\subsection*{B.3 Adapter Bootstrap Protocol}\label{b.3-adapter-bootstrap-protocol}

\texttt{spl3/adapters/\_\_init\_\_.py} resolves adapters in three-step priority order:

\begin{enumerate}
\def\labelenumi{\arabic{enumi}.}
\tightlist
\item
  \textbf{dd-llm bridge} --- if installed, registered first for all cloud model prefixes.
\item
  \textbf{Bespoke fallbacks} --- \texttt{claude\_cli} and \texttt{openrouter} registered as standalone adapters for environments without dd-llm.
\item
  \textbf{Always-available} --- \texttt{ollama} and \texttt{momagrid} always registered last as unconditional fallbacks.
\end{enumerate}

\subsection*{B.4 Two-Method Interface}\label{b.4-two-method-interface}

All adapters implement:

\begin{verbatim}
class BaseAdapter:
    def generate(
        self, prompt: str, system: str, model: str, **kw
    ) -> GenerationResult: ...

    def generate_multimodal(
        self, content: list[dict], system: str, model: str, **kw
    ) -> GenerationResult: ...
\end{verbatim}

The executor dispatches to \texttt{generate\_multimodal()} when any \texttt{INPUT} variable carries an \texttt{IMAGE}, \texttt{AUDIO}, or \texttt{VIDEO} type annotation; otherwise \texttt{generate()}. \texttt{GenerationResult} carries \texttt{content:\ str}, \texttt{model:\ str}, \texttt{latency\_ms:\ float}, and \texttt{token\_counts:\ dict}.

\subsection*{B.5 Momagrid Response Protocol}\label{b.5-momagrid-response-protocol}

The Momagrid hub returns a nested result object that must not be flattened:

\begin{verbatim}
{
  "state": "COMPLETE",
  "result": {
    "content": "...",
    "latency_ms": 342,
    "agent_name": "worker-node-07"
  }
}
\end{verbatim}

\texttt{agent\_name} is load-bearing for multi-node routing and audit logging.

\section*{Appendix C: Verifier Ladder --- Setup and Configuration}\label{appendix-c-verifier-ladder-setup-and-configuration}

\subsection*{C.1 Rung R1: SymPy (zero additional setup)}\label{c.1-rung-r1-sympy-zero-additional-setup}

SymPy ships with \texttt{spl-llm}:

\begin{verbatim}
pip install spl-llm          # SymPy included
spl3 run workflow.spl --kernel --llm ollama:gemma3
\end{verbatim}

The \texttt{-\/-kernel} flag starts an IPython kernel with \texttt{python3} kernelspec. SymPy operations are registered as SPL tools in \texttt{cookbook/67\_symbolic\_math/tools.py} and are available to any workflow that imports that file or registers equivalent tools.

\subsection*{C.2 Rung R2: SageMath}\label{c.2-rung-r2-sagemath}

SageMath requires a separate install due to its size (\textasciitilde1 GB):

\begin{verbatim}
# Option A — conda (recommended)
conda install -c conda-forge sagemath

# Option B — system package (Ubuntu/Debian)
sudo apt install sagemath

# Register the SageMath Jupyter kernelspec
sage -python -m sage.repl.ipython_kernel.install

# Run with SageMath kernel
spl3 run workflow.spl --kernel --kernel-name sagemath --llm ollama:gemma3
\end{verbatim}

SPL discovers the kernel via \texttt{ensure\_kernelspec("sagemath")} in \texttt{spl3/kernel.py}. The same \texttt{SOLVE}/\texttt{ASSERT} primitives work unchanged; only the execution environment differs.

\subsection*{C.3 Rung R3: Lean 4 with Lean REPL}\label{c.3-rung-r3-lean-4-with-lean-repl}

Lean 4 and the REPL are installed via the one-shot setup script:

\begin{verbatim}
# Full install: elan + Lean 4 + leanprover-community/repl (pinned v4.30.0)
bash cookbook/tools/lean/setup_lean.sh

# Optionally with Mathlib (~2 GB download, required for library search)
bash cookbook/tools/lean/setup_lean.sh --with-mathlib
\end{verbatim}

The script installs \texttt{elan} (Lean version manager) to \texttt{\$ELAN\_HOME} (default \texttt{/opt/lean}) and builds the REPL with \texttt{lake}. The SPL Lean bridge (\texttt{spl3/lean\_bridge.py}) wraps the REPL in a persistent session:

\begin{verbatim}
from spl3.lean_bridge import LeanREPL

repl = LeanREPL().start()           # stdlib only
repl = LeanREPL.mathlib().start()   # with Mathlib imports
\end{verbatim}

The bridge passes 15/15 unit tests in \texttt{tests/test\_lean\_bridge.py}, covering statement checking, proof verification, \texttt{exact?} citation, timeout recovery, and REPL restart.

\begin{verbatim}
# Verify installation
pytest tests/test_lean_bridge.py -v

# Run recipe 76 (the full proof pipeline)
spl3 run cookbook/76_lean_proof/lean_proof.spl --kernel --llm claude_cli
\end{verbatim}

\subsection*{C.4 Environment Variables}\label{c.4-environment-variables}

\begin{longtable}[]{@{}lll@{}}
\toprule\noalign{}
Variable & Default & Purpose \\
\midrule\noalign{}
\endhead
\bottomrule\noalign{}
\endlastfoot
\texttt{MOMAGRID\_HUB\_URL} & \texttt{http://localhost:9000} & Momagrid hub endpoint \\
\texttt{OPENROUTER\_API\_KEY} & --- & OpenRouter authentication \\
\texttt{ELAN\_HOME} & \texttt{/opt/lean} & Lean toolchain root \\
\texttt{SPL\_LEAN\_REPL\_PATH} & auto-detected & Path to \texttt{lean\_repl} binary \\
\texttt{SPL\_KERNEL\_TIMEOUT} & \texttt{120} & Kernel execution timeout (seconds) \\
\texttt{SPL\_MAX\_ITERATIONS} & \texttt{15} & Default WHILE loop guard \\
\end{longtable}

\begin{quote}
\textbf{Note on \texttt{SPL\_MAX\_ITERATIONS}:} Complex workflows with multiple refinement loops (e.g., formalization + proof repair in Lean, or multi-step verification chains) may require higher iteration limits. If you encounter \texttt{MaxIterationsReached} exceptions, increase this value (e.g., \texttt{export\ SPL\_MAX\_ITERATIONS=25}) to allow nested loops to complete without premature exit. The default of 15 balances safety against infinite loops with flexibility for most use cases.
\end{quote}

\section*{Appendix D: SPL Cookbook Recipe Catalog}\label{appendix-d-spl-cookbook-recipe-catalog}

The cookbook contains 78 recipes (00--77) organized across 8 categories, developed iteratively as the project evolved. Each recipe demonstrates a specific pattern or construct. Active recipes run in the batch runner (\texttt{python\ cookbook/run\_all.py}).

\textbf{Purpose:} This recipe collection serves two roles. (1) \textbf{Feature verification}: each recipe validates a distinct SPL language construct or adapter capability, enabling rapid regression testing across new runtime versions and model providers. (2) \textbf{Code-RAG training data}: the description/source pairs are indexed and retrieved by the \texttt{text2spl} tool to guide LLM-based code generation, improving quality through in-domain examples. Community contributions are welcomed; recipes that add new constructs or clarify existing patterns strengthen both validation and the retrieval index. A decentralized \textbf{workflow registry} is planned for future work, allowing teams to publish vetted recipes and share standard patterns across organizations.

\subsection*{D.1 Foundations (00--09)}\label{d.1-foundations-0009}

\begin{longtable}[]{@{}
  >{\raggedright\arraybackslash}p{0.5cm}
  >{\raggedright\arraybackslash}p{2.6cm}
  >{\raggedright\arraybackslash}p{9.9cm}@{}}
\toprule\noalign{}
\begin{minipage}[b]{\linewidth}\raggedright
ID
\end{minipage} & \begin{minipage}[b]{\linewidth}\raggedright
Name
\end{minipage} & \begin{minipage}[b]{\linewidth}\raggedright
Description
\end{minipage} \\
\midrule\noalign{}
\endhead
\bottomrule\noalign{}
\endlastfoot
00 & Recipe Maker & Auto-generate SPL recipes from natural language specs \\
01 & Hello World & Basic PROMPT and WORKFLOW: single LLM call \\
02 & Ollama Proxy & Route requests through Ollama; test local model access \\
03 & Multilingual & Parallel LLM calls in multiple languages; vote on result \\
04 & Model Showdown & A/B comparison of two models on identical task \\
05 & Self-Refine & WHILE loop refinement until quality gate passes \\
06 & ReAct Agent & Reasoning + acting: LLM thinks, then calls tools \\
07 & Safe Generation & Content filtering and safety guardrails \\
08 & RAG Query & Retrieval-augmented generation; query external knowledge base \\
09 & Chain-of-Thought & Multi-step reasoning with intermediate checkpoints \\
\end{longtable}

\subsection*{D.2 Content Generation (10--19)}\label{d.2-content-generation-1019}

\begin{longtable}[]{@{}
  >{\raggedright\arraybackslash}p{0.5cm}
  >{\raggedright\arraybackslash}p{2.6cm}
  >{\raggedright\arraybackslash}p{9.9cm}@{}}
\toprule\noalign{}
\begin{minipage}[b]{\linewidth}\raggedright
ID
\end{minipage} & \begin{minipage}[b]{\linewidth}\raggedright
Name
\end{minipage} & \begin{minipage}[b]{\linewidth}\raggedright
Description
\end{minipage} \\
\midrule\noalign{}
\endhead
\bottomrule\noalign{}
\endlastfoot
10 & Batch Test & Parallel inference on a batch of prompts; collect and compare \\
11 & Debate Arena & Two agents debate a topic; third agent judges \\
12 & Plan and Execute & LLM decomposes task into steps; executor validates each \\
13 & Map-Reduce & Map computation across items; reduce results with aggregation \\
14 & Multi-Agent & Choreograph multiple specialized agents toward shared goal \\
15 & Code Review & LLM reviews code; iterative feedback loop until approved \\
16 & Reflection & Generate response; LLM reflects on quality; iterate if needed \\
17 & Tree of Thought & Explore multiple reasoning paths; prune low-confidence branches \\
18 & Guardrails & Runtime validation of LLM output against constraints \\
19 & Memory Conversation & Multi-turn dialogue with persistent context/memory buffer \\
\end{longtable}

\subsection*{D.3 Analysis and Extraction (20--29)}\label{d.3-analysis-and-extraction-2029}

\begin{longtable}[]{@{}
  >{\raggedright\arraybackslash}p{0.5cm}
  >{\raggedright\arraybackslash}p{2.6cm}
  >{\raggedright\arraybackslash}p{9.9cm}@{}}
\toprule\noalign{}
\begin{minipage}[b]{\linewidth}\raggedright
ID
\end{minipage} & \begin{minipage}[b]{\linewidth}\raggedright
Name
\end{minipage} & \begin{minipage}[b]{\linewidth}\raggedright
Description
\end{minipage} \\
\midrule\noalign{}
\endhead
\bottomrule\noalign{}
\endlastfoot
20 & Ensemble Voting & Run multiple models; majority vote on classification \\
21 & Multi-Model Pipeline & Sequence models: model A → model B, pass-through \\
22 & Text2SPL Demo & Natural language → SPL source code generation \\
23 & Structured Output & Enforce JSON/XML schema on LLM response; validate format \\
24 & Few-Shot Prompting & In-context learning: embed examples in PROMPT \\
25 & Nested Procedures & Call workflows from within workflows; composition \\
26 & A/B Test & Run same task on two variants; measure performance delta \\
27 & Data Extraction & Extract structured fields from unstructured text \\
28 & Support Triage & Classify support tickets; route to specialized handlers \\
29 & Meeting Actions & Extract action items, attendees, dates from meeting transcripts \\
\end{longtable}

\subsection*{D.4 Education and Tutoring (30--39)}\label{d.4-education-and-tutoring-3039}

\begin{longtable}[]{@{}
  >{\raggedright\arraybackslash}p{0.5cm}
  >{\raggedright\arraybackslash}p{2.6cm}
  >{\raggedright\arraybackslash}p{9.9cm}@{}}
\toprule\noalign{}
\begin{minipage}[b]{\linewidth}\raggedright
ID
\end{minipage} & \begin{minipage}[b]{\linewidth}\raggedright
Name
\end{minipage} & \begin{minipage}[b]{\linewidth}\raggedright
Description
\end{minipage} \\
\midrule\noalign{}
\endhead
\bottomrule\noalign{}
\endlastfoot
30 & Code Generator & Generate code from natural language specification \\
31 & Sentiment Pipeline & Analyze text sentiment; classify as positive/negative/neutral \\
32 & Socratic Tutor & Ask clarifying questions; guide learner to answer \\
33 & Interview Simulator & Conduct mock interview; score responses in real-time \\
34 & Progressive Summarizer & Iteratively summarize text to target length/complexity \\
35 & Hypothesis Tester & LLM proposes hypothesis; test against data; iterate \\
36 & Tool-Use & Register and call custom Python functions as tools \\
37 & Headline News Aggregator & Fetch, summarize, and rank news headlines by relevance \\
38 & Bedrock Quickstart & AWS Bedrock adapter: example integration \\
39 & Vertex AI Quickstart & Google Vertex AI adapter: example integration \\
\end{longtable}

\subsection*{D.5 Cloud and Infrastructure (40--49)}\label{d.5-cloud-and-infrastructure-4049}

\begin{longtable}[]{@{}
  >{\raggedright\arraybackslash}p{0.5cm}
  >{\raggedright\arraybackslash}p{2.6cm}
  >{\raggedright\arraybackslash}p{9.9cm}@{}}
\toprule\noalign{}
\begin{minipage}[b]{\linewidth}\raggedright
ID
\end{minipage} & \begin{minipage}[b]{\linewidth}\raggedright
Name
\end{minipage} & \begin{minipage}[b]{\linewidth}\raggedright
Description
\end{minipage} \\
\midrule\noalign{}
\endhead
\bottomrule\noalign{}
\endlastfoot
40 & Azure OpenAI Quickstart & Azure OpenAI adapter: API key config and request routing \\
41 & Human Steering & Pause for user input; incorporate feedback before proceeding \\
42 & Knowledge Synthesis & Combine insights from multiple documents into unified summary \\
43 & Prompt Self-Tuning & Automatically optimize prompt structure based on output quality \\
44 & Adaptive Failover & Primary model fails → fallback to secondary model \\
45 & Vision to Action & Process image input; classify scene; trigger corresponding action \\
47 & arXiv Morning Brief & Fetch latest papers in topic; summarize key findings \\
48 & Credit Risk Assessment & Evaluate financial data; output risk score with explanation \\
49 & Regulatory News Audit & Monitor regulatory news; flag relevant updates for compliance team \\
\end{longtable}

\subsection*{D.6 Multimodal (50--59)}\label{d.6-multimodal-5059}

\begin{longtable}[]{@{}
  >{\raggedright\arraybackslash}p{0.5cm}
  >{\raggedright\arraybackslash}p{2.6cm}
  >{\raggedright\arraybackslash}p{9.9cm}@{}}
\toprule\noalign{}
\begin{minipage}[b]{\linewidth}\raggedright
ID
\end{minipage} & \begin{minipage}[b]{\linewidth}\raggedright
Name
\end{minipage} & \begin{minipage}[b]{\linewidth}\raggedright
Description
\end{minipage} \\
\midrule\noalign{}
\endhead
\bottomrule\noalign{}
\endlastfoot
50 & Code Pipeline & Multimodal: parse code images → OCR → execute \\
51 & Image Caption & Generate descriptive captions for images \\
52 & Audio Summary & Transcribe audio; summarize transcript \\
53 & Video Summary & Extract keyframes; caption each; synthesize summary \\
54 & Text to Image & Generate image from text prompt (DALL-E, Stable Diffusion) \\
55 & Text to Speech & Convert text output to audio/voice \\
56 & Text to Video & Generate video from text script (experimental) \\
57 & Image Format Conversion & Convert image between formats (PNG $\leftrightarrow$ JPEG $\leftrightarrow$ WebP) \\
58 & Image Restyle & Apply style transfer or filter to image \\
59 & Audio Format Conversion & Convert audio between codecs (MP3 $\leftrightarrow$ WAV $\leftrightarrow$ M4A) \\
\end{longtable}

\subsection*{D.7 Parallel and Compilation (60--69)}\label{d.7-parallel-and-compilation-6069}

\begin{longtable}[]{@{}
  >{\raggedright\arraybackslash}p{0.5cm}
  >{\raggedright\arraybackslash}p{2.6cm}
  >{\raggedright\arraybackslash}p{9.9cm}@{}}
\toprule\noalign{}
\begin{minipage}[b]{\linewidth}\raggedright
ID
\end{minipage} & \begin{minipage}[b]{\linewidth}\raggedright
Name
\end{minipage} & \begin{minipage}[b]{\linewidth}\raggedright
Description
\end{minipage} \\
\midrule\noalign{}
\endhead
\bottomrule\noalign{}
\endlastfoot
60 & Voice Dialogue & Real-time speech $\leftrightarrow$ text; interactive conversation \\
61 & Video to Audio & Extract audio track from video \\
62 & Video to Image & Extract keyframes or create GIF from video \\
63 & Parallel Code Review & Review multiple code files concurrently via CALL PARALLEL \\
64 & Parallel News Digest & Fetch and summarize multiple articles in parallel \\
65 & LLM Splc & Compile \texttt{.spl} to imperative code (Go/TypeScript) \\
66 & Stock Analysis & Fetch price data; LLM generates trading signals with explanation \\
67 & Symbolic Math & SymPy backend: algebraic problem solving (instances verified) \\
68 & Problem Generator & Auto-generate educational problems from topic + difficulty \\
69 & Notebook Generator & Convert \texttt{.spl} workflow to Jupyter notebook \\
\end{longtable}

\subsection*{D.8 Verifier Ladder and Concept-Book (70--77)}\label{d.8-verifier-ladder-and-concept-book-7077}

\begin{longtable}[]{@{}
  >{\raggedright\arraybackslash}p{0.5cm}
  >{\raggedright\arraybackslash}p{2.6cm}
  >{\raggedright\arraybackslash}p{9.9cm}@{}}
\toprule\noalign{}
\begin{minipage}[b]{\linewidth}\raggedright
ID
\end{minipage} & \begin{minipage}[b]{\linewidth}\raggedright
Name
\end{minipage} & \begin{minipage}[b]{\linewidth}\raggedright
Description
\end{minipage} \\
\midrule\noalign{}
\endhead
\bottomrule\noalign{}
\endlastfoot
70 & Linear Algebra Core Concepts & Build concept graph for linalg topics; trace prerequisites \\
71 & Linear Algebra Concept-Book & Interactive HTML learning path: definitions → theorems → proofs \\
72 & Verify arXiv References & LLM cites paper; kernel verifies bibliographic accuracy \\
73 & Intro Geometry Concept-Book & Concept-book for geometry: points → lines → shapes → proofs \\
74 & Generic Concept-Book & Template for concept-books on any STEM topic \\
75 & SageMath Solver & Sage kernel: instances SymPy cannot reach (Galois, eigenvalues, etc.) \\
76 & Lean Proof Verifier & Lean 4 + mathlib: prove statements; kernel checks proof \\
77 & \textbf{Neurosymbolic Solver} & \textbf{Main experiment: unified workflow across SymPy → Sage → Lean rungs} \\
\end{longtable}

\section*{Appendix E: Experiment --- Full Results and Methodology}\label{appendix-e-experiment-full-results-and-methodology}

\subsection*{E.1 Experimental Protocol}\label{e.1-experimental-protocol}

Two sessions were run on the same local workstation (Ollama for 9 local models + \texttt{claude\_cli} for sonnet-4-6):

\begin{longtable}[]{@{}llll@{}}
\toprule\noalign{}
Session & Cells & Repeats & ID \\
\midrule\noalign{}
\endhead
\bottomrule\noalign{}
\endlastfoot
Pilot & 400 & r=1 & \texttt{exp-20260615-073849} \\
Repeated & 1200 & r=3 & \texttt{exp-20260615-191224} \\
\end{longtable}

\textbf{Pass oracle:} SPL status codes --- solver arm: \texttt{complete} (chain kernel-verified); LLM-only arm: \texttt{complete} or \texttt{unverified\_success} (non-empty response)\\
\textbf{Database:} \texttt{cookbook/77\_neurosymbolic/experiment\_results.db} (SQLite)

The dual-arm design uses a single \texttt{.spl} workflow (\texttt{symbolic\_math.spl}) with \texttt{enable\_solver} as a runtime parameter. The harness (\texttt{run\_experiment.py}) iterates all (model $\times$ problem $\times$ solver\_mode) cells sequentially, invoking \texttt{spl3\ run} for each and persisting structured results to the DB. Status codes reported in the DB:

\begin{itemize}
\tightlist
\item
  \texttt{complete} --- full chain verified by the symbolic kernel (solver arm) or non-empty LLM response (LLM-only arm)
\item
  \texttt{unverified\_success} --- LLM-only arm produced output but with steps=0
\item
  \texttt{solver\_error} --- kernel rejected one or more steps (expression evaluation failed)
\item
  \texttt{plan\_error} --- plan parsed but semantically invalid (e.g., wrong step count)
\end{itemize}

\subsection*{E.2 Problem Set}\label{e.2-problem-set}

The 20 problems span six tiers across two backends. T0--T2 use SymPy (10 problems); T3--T5 use SageMath (10 problems). Each problem has a known ground-truth answer available in the kernel but not used as the pass oracle (status codes provide it).

\begin{longtable}[]{@{}
  >{\raggedright\arraybackslash}p{0.6cm}
  >{\raggedright\arraybackslash}p{1cm}
  >{\raggedright\arraybackslash}p{0.6cm}
  >{\raggedright\arraybackslash}p{5.0cm}
  >{\raggedright\arraybackslash}p{5.5cm}@{}}
\toprule\noalign{}
\begin{minipage}[b]{\linewidth}\raggedright
Tier
\end{minipage} & \begin{minipage}[b]{\linewidth}\raggedright
Backend
\end{minipage} & \begin{minipage}[b]{\linewidth}\raggedright
Count
\end{minipage} & \begin{minipage}[b]{\linewidth}\raggedright
Category
\end{minipage} & \begin{minipage}[b]{\linewidth}\raggedright
Examples
\end{minipage} \\
\midrule\noalign{}
\endhead
\bottomrule\noalign{}
\endlastfoot
T0 & SymPy & 2 & Poly single-step & $d/dx(x^4 - 2x^2 + 1)$; simplify$(x^2-1)/(x-1)$ \\
T1 & SymPy & 4 & Poly multi-step & expand$(x+1)^2 \to$ factor; diff $3x^3-x \to$ factor $\to$ solve \\
T2 & SymPy & 4 & Transcendental/limits/series & $\sin(x)/x \to 0$; Taylor $\sin(x)$ $n=5$ \\
T3 & Sage & 4 & Integration/systems/eigenvalues & $\int\!\sqrt{4-x^2}\,dx$; $x+y=5, x-y=1$; eig{[}{[}1,2{]},{[}3,4{]}{]} \\
T4 & Sage & 4 & Laplace/ODEs/sums/roots & $\mathcal{L}\{e^{-2t}\}$; $y'=y, y(0)=1$; $\sum 1/n^2$; roots $x^4-1$ \\
T5 & Sage & 2 & 2nd-order ODE + verify & $y''-3y'+2y=0$; $\mathcal{L}^{-1}\{s/(s^2+4)\}$ verify \\
\end{longtable}

\subsection*{\texorpdfstring{E.3 Per-Tier Pass Rates (r=3 session, \texttt{exp-20260615-191224})}{E.3 Per-Tier Pass Rates (r=3 session, exp-20260615-191224)}}\label{e.3-per-tier-pass-rates-r3-session-exp-20260615-191224}

\textbf{Solver arm pass rate by model and tier (\%) --- mean over 3 runs/cell:}

\begin{longtable}[]{@{}
  >{\raggedright\arraybackslash}p{(\columnwidth - 14\tabcolsep) * \real{0.20}}
  >{\raggedright\arraybackslash}p{(\columnwidth - 14\tabcolsep) * \real{0.09}}
  >{\raggedright\arraybackslash}p{(\columnwidth - 14\tabcolsep) * \real{0.09}}
  >{\raggedright\arraybackslash}p{(\columnwidth - 14\tabcolsep) * \real{0.09}}
  >{\raggedright\arraybackslash}p{(\columnwidth - 14\tabcolsep) * \real{0.09}}
  >{\raggedright\arraybackslash}p{(\columnwidth - 14\tabcolsep) * \real{0.09}}
  >{\raggedright\arraybackslash}p{(\columnwidth - 14\tabcolsep) * \real{0.09}}
  >{\raggedright\arraybackslash}p{(\columnwidth - 14\tabcolsep) * \real{0.09}}@{}}
\toprule\noalign{}
\begin{minipage}[b]{\linewidth}\raggedright
Model
\end{minipage} & \begin{minipage}[b]{\linewidth}\raggedright
T0
\end{minipage} & \begin{minipage}[b]{\linewidth}\raggedright
T1
\end{minipage} & \begin{minipage}[b]{\linewidth}\raggedright
T2
\end{minipage} & \begin{minipage}[b]{\linewidth}\raggedright
T3
\end{minipage} & \begin{minipage}[b]{\linewidth}\raggedright
T4
\end{minipage} & \begin{minipage}[b]{\linewidth}\raggedright
T5
\end{minipage} & \begin{minipage}[b]{\linewidth}\raggedright
Overall
\end{minipage} \\
\midrule\noalign{}
\endhead
\bottomrule\noalign{}
\endlastfoot
gemma4:e2b & 100 & 100 & 100 & 100 & 75 & 83 & 93 \\
sonnet-4-6 & 100 & 92 & 100 & 100 & 58 & 50 & 85 \\
rnj-1 & 83 & 100 & 100 & 50 & 92 & 50 & 82 \\
gemma3 & 83 & 100 & 75 & 75 & 50 & 50 & 73 \\
qwen2.5 & 100 & 100 & 92 & 67 & 33 & 33 & 72 \\
phi4 & 67 & 75 & 100 & 50 & 50 & 50 & 67 \\
llama3.2 & 67 & 83 & 100 & 42 & 42 & 50 & 65 \\
deepseek-v2:16b & 50 & 75 & 50 & 58 & 42 & 33 & 53 \\
lfm2.5 & 67 & 33 & 25 & 50 & 25 & 33 & 37 \\
phi3 & 33 & 33 & 33 & 42 & 33 & 0 & 32 \\
\textbf{Tier avg} & \textbf{75} & \textbf{79} & \textbf{77} & \textbf{63} & \textbf{50} & \textbf{43} & --- \\
\end{longtable}

\textbf{LLM-only arm pass rate by model and tier (\%) --- mean over 3 runs/cell:}

\begin{longtable}[]{@{}llllllll@{}}
\toprule\noalign{}
Model & T0 & T1 & T2 & T3 & T4 & T5 & Overall \\
\midrule\noalign{}
\endhead
\bottomrule\noalign{}
\endlastfoot
sonnet-4-6 & 100 & 100 & 100 & 100 & 100 & 100 & 100 \\
rnj-1 & 100 & 100 & 100 & 100 & 100 & 100 & 100 \\
qwen2.5 & 100 & 100 & 100 & 100 & 100 & 100 & 100 \\
phi4 & 100 & 100 & 100 & 100 & 100 & 100 & 100 \\
phi3 & 100 & 100 & 100 & 100 & 100 & 100 & 100 \\
gemma3 & 100 & 100 & 100 & 100 & 100 & 100 & 100 \\
llama3.2 & 100 & 100 & 100 & 100 & 100 & 100 & 100 \\
deepseek-v2:16b & 100 & 100 & 100 & 100 & 100 & 100 & 100 \\
gemma4:e2b & 100 & 100 & 100 & 92 & 92 & 100 & 97 \\
lfm2.5 & 100 & 67 & 75 & 92 & 75 & 50 & 77 \\
\end{longtable}

\subsection*{E.4 Failure Mode Breakdown (Solver Arm, r=3, 60 runs per model)}\label{e.4-failure-mode-breakdown-solver-arm-r3-60-runs-per-model}

\begin{longtable}[]{@{}
  >{\raggedright\arraybackslash}p{(\columnwidth - 10\tabcolsep) * \real{0.25}}
  >{\raggedright\arraybackslash}p{(\columnwidth - 10\tabcolsep) * \real{0.15}}
  >{\raggedright\arraybackslash}p{(\columnwidth - 10\tabcolsep) * \real{0.15}}
  >{\raggedright\arraybackslash}p{(\columnwidth - 10\tabcolsep) * \real{0.15}}
  >{\raggedright\arraybackslash}p{(\columnwidth - 10\tabcolsep) * \real{0.15}}
  >{\raggedright\arraybackslash}p{(\columnwidth - 10\tabcolsep) * \real{0.15}}@{}}
\toprule\noalign{}
\begin{minipage}[b]{\linewidth}\raggedright
Model
\end{minipage} & \begin{minipage}[b]{\linewidth}\raggedright
\texttt{complete}
\end{minipage} & \begin{minipage}[b]{\linewidth}\raggedright
\texttt{solver\_error}
\end{minipage} & \begin{minipage}[b]{\linewidth}\raggedright
\texttt{plan\_error}
\end{minipage} & \begin{minipage}[b]{\linewidth}\raggedright
\texttt{other}
\end{minipage} & \begin{minipage}[b]{\linewidth}\raggedright
\texttt{runs}
\end{minipage} \\
\midrule\noalign{}
\endhead
\bottomrule\noalign{}
\endlastfoot
gemma4:e2b & 56 & 0 & 4 & 0 & 60 \\
sonnet-4-6 & 51 & 8 & 0 & 1 & 60 \\
rnj-1 & 49 & 11 & 0 & 0 & 60 \\
gemma3 & 44 & 16 & 0 & 0 & 60 \\
qwen2.5 & 43 & 17 & 0 & 0 & 60 \\
phi4 & 40 & 20 & 0 & 0 & 60 \\
llama3.2 & 39 & 21 & 0 & 0 & 60 \\
deepseek-v2:16b & 32 & 28 & 0 & 0 & 60 \\
lfm2.5 & 22 & 11 & 26 & 1 & 60 \\
phi3 & 19 & 41 & 0 & 0 & 60 \\
\end{longtable}

\texttt{plan\_format\_error} is zero across all models and both sessions. The \texttt{other} column captures rare edge cases: \texttt{sonnet-4-6} has one \texttt{unknown} status (p012/T1, run 1) and \texttt{lfm2.5} has one \texttt{narration\_error} --- neither is a \texttt{solver\_error} or \texttt{plan\_error}. \texttt{lfm2.5} is the sole \texttt{plan\_error} outlier (26/60), indicating a plan-generation failure distinct from kernel expression errors. \texttt{phi3} accumulates 41 \texttt{solver\_error} failures across 60 runs, the most of any model, reflecting its difficulty naming correct Sage operations.

\subsection*{E.5 r=1 vs r=3 Comparison (Pilot vs Repeated Run)}\label{e.5-r1-vs-r3-comparison-pilot-vs-repeated-run}

The pilot session (\texttt{exp-20260615-073849}, r=1) and repeated session (\texttt{exp-20260615-191224}, r=3) ran the identical 10-model $\times$ 20-problem $\times$ 2-arm design. Solver arm pass rates (overall \%):

\begin{longtable}[]{@{}llll@{}}
\toprule\noalign{}
Model & r=1 (pilot) & r=3 (repeated) & $\Delta$ \\
\midrule\noalign{}
\endhead
\bottomrule\noalign{}
\endlastfoot
gemma4:e2b & 95 & 93 & -2 \\
sonnet-4-6 & 85 & 85 & 0 \\
rnj-1 & 90 & 82 & -8 \\
gemma3 & 70 & 73 & +3 \\
qwen2.5 & 75 & 72 & -3 \\
\textbf{phi4} & \textbf{85} & \textbf{67} & \textbf{-18} \\
llama3.2 & 65 & 65 & 0 \\
deepseek-v2:16b & 45 & 53 & +8 \\
lfm2.5 & 45 & 37 & -8 \\
phi3 & 30 & 32 & +2 \\
\end{longtable}

Top-3 and bottom-2 rankings are stable. The largest swing is phi4 (-18 pp), revealing pilot over-estimation. \texttt{sonnet-4-6} and \texttt{llama3.2} are perfectly stable at 85\% and 65\% respectively, showing consistent solver behavior for those models. The r=3 session is the authoritative result used in \S\ref{empirical-evaluation}.

\subsection*{E.6 Judge Prompt}\label{e.6-judge-prompt}

The LLM sanity gate uses the following prompt for each run:

\begin{verbatim}
You are a mathematics judge. You will be given a problem and a proposed answer.
Determine whether the answer is mathematically correct.

Problem: {problem}
Proposed answer: {answer}

Reply with exactly one word: "pass" if the answer is correct, "fail" if it is not.
Do not explain. Do not add punctuation.
\end{verbatim}

The judge model (\texttt{claude-sonnet-4-6}) is run independently of the experiment sessions and has no access to the SymPy ground truth. \texttt{qwen3} results are excluded from all tables: root-cause analysis found that \texttt{qwen3.5:9b} runs in extended thinking mode by default, exhausting its token budget on internal deliberation before emitting any structured output. This is a model-interface incompatibility, not a capability failure --- the thinking trace shows correct reasoning. The re-run will use a non-thinking \texttt{qwen3} variant or disable thinking mode via the \texttt{/no\_think} flag.

\section*{Appendix F: Compilation Pipeline --- Target Examples}\label{appendix-f-compilation-pipeline-target-examples}

The \texttt{spl3\ splc} compiler translates a \texttt{.spl} workflow to idiomatic code in each target framework. All examples below compile from the same source: a minimal two-mode workflow that decomposes a math problem and verifies it with SymPy.

\subsection*{\texorpdfstring{Source: \texttt{verify\_step.spl}}{Source: verify\_step.spl}}\label{source-verify_step.spl}

\begin{verbatim}
WORKFLOW verify_step
  INPUT: @problem TEXT
  OUTPUT: @report TEXT
DO
  GENERATE decompose(@problem) INTO @steps
    SELECT @problem AS problem
    PROMPT "Decompose into expr|op steps, one per line: {@problem}"

  SOLVE @result SYMPY := "solve_chain({@steps})"

  ASSERT verify({@result}, {@problem})
    OTHERWISE RETURN @report WITH status = 'verification_failed'

  GENERATE narrate(@result, @problem) INTO @report
    SELECT @result AS result, @problem AS problem
    PROMPT "Explain this verified result in plain language: {@result}"

  RETURN @report WITH status = 'complete'
END
\end{verbatim}

\subsection*{F.1 LangGraph Target}\label{f.1-langgraph-target}

\begin{verbatim}
spl3 splc verify_step.spl --target langgraph
\end{verbatim}

\begin{verbatim}
# verify_step_langgraph.py  — generated by spl3 splc
from langgraph.graph import StateGraph, END
from typing import TypedDict

class State(TypedDict):
    problem: str
    steps: str
    result: str
    report: str
    status: str

def node_decompose(state: State) -> State:
    prompt = f"Decompose into expr|op steps, one per line: {state['problem']}"
    state["steps"] = llm_call(prompt)
    return state

def node_solve(state: State) -> State:
    state["result"] = kernel_exec(f"solve_chain({state['steps']!r})")
    return state

def node_assert(state: State) -> State:
    ok = kernel_exec(f"bool(verify({state['result']!r}, {state['problem']!r}))")
    state["status"] = "ok" if ok == "True" else "verification_failed"
    return state

def node_narrate(state: State) -> State:
    prompt = f"Explain this verified result in plain language: {state['result']}"
    state["report"] = llm_call(prompt)
    return state

def route_assert(state: State) -> str:
    return "narrate" if state["status"] == "ok" else END

graph = StateGraph(State)
graph.add_node("decompose", node_decompose)
graph.add_node("solve", node_solve)
graph.add_node("assert_gate", node_assert)
graph.add_node("narrate", node_narrate)
graph.set_entry_point("decompose")
graph.add_edge("decompose", "solve")
graph.add_edge("solve", "assert_gate")
graph.add_conditional_edges("assert_gate", route_assert)
graph.add_edge("narrate", END)
app = graph.compile()
\end{verbatim}

\subsection*{F.2 Go Target}\label{f.2-go-target}

\begin{verbatim}
spl3 splc verify_step.spl --target go
\end{verbatim}

{\footnotesize
\begin{verbatim}
// verify_step.go — generated by spl3 splc
package main

import (
    "fmt"
    "splruntime"   // SPL Go runtime shim
)

type VerifyStepState struct {
    Problem string
    Steps   string
    Result  string
    Report  string
    Status  string
}

func VerifyStep(problem string) (string, error) {
    s := &VerifyStepState{Problem: problem}

    var err error
    s.Steps, err = splruntime.Generate("Decompose into expr|op steps, one per line: "+s.Problem)
    if err != nil { return "", err }

    s.Result, err = splruntime.KernelExec(fmt.Sprintf("solve_chain(%q)", s.Steps))
    if err != nil { return "", err }

    ok, err := splruntime.KernelAssert(
        fmt.Sprintf("bool(verify(%q, %q))", s.Result, s.Problem))
    if err != nil || !ok {
        return "", splruntime.CommitStatus("verification_failed")
    }

    s.Report, err = splruntime.Generate(
        "Explain this verified result in plain language: " + s.Result)
    if err != nil { return "", err }

    return s.Report, nil
}
\end{verbatim}
}

\subsection*{F.3 TypeScript Target}\label{f.3-typescript-target}

\begin{verbatim}
spl3 splc verify_step.spl --target typescript
\end{verbatim}

\begin{verbatim}
// verifyStep.ts — generated by spl3 splc
import { generate, kernelExec, kernelAssert } from "./spl-runtime";

interface VerifyStepState {
  problem: string;
  steps?: string;
  result?: string;
  report?: string;
}

export async function verifyStep(problem: string): Promise<string> {
  const s: VerifyStepState = { problem };

  s.steps = await generate(
    `Decompose into expr|op steps, one per line: ${s.problem}`
  );

  s.result = await kernelExec(`solve_chain(${JSON.stringify(s.steps)})`);

  const ok = await kernelAssert(
    `bool(verify(${JSON.stringify(s.result)}, ${JSON.stringify(s.problem)}))`
  );
  if (!ok) throw new Error("verification_failed");

  s.report = await generate(
    `Explain this verified result in plain language: ${s.result}`
  );
  return s.report;
}
\end{verbatim}

\subsection*{F.4 DODA Invariant Across Targets}\label{f.4-doda-invariant-across-targets}

The \texttt{.spl} source is unchanged across all three targets. The compiler manages:

\begin{longtable}[]{@{}
  >{\raggedright\arraybackslash}p{(\columnwidth - 6\tabcolsep) * \real{0.18}}
  >{\raggedright\arraybackslash}p{(\columnwidth - 6\tabcolsep) * \real{0.20}}
  >{\raggedright\arraybackslash}p{(\columnwidth - 6\tabcolsep) * \real{0.30}}
  >{\raggedright\arraybackslash}p{(\columnwidth - 6\tabcolsep) * \real{0.3}}@{}}
\toprule\noalign{}
\begin{minipage}[b]{\linewidth}\raggedright
SPL construct
\end{minipage} & \begin{minipage}[b]{\linewidth}\raggedright
LangGraph
\end{minipage} & \begin{minipage}[b]{\linewidth}\raggedright
Go
\end{minipage} & \begin{minipage}[b]{\linewidth}\raggedright
TypeScript
\end{minipage} \\
\midrule\noalign{}
\endhead
\bottomrule\noalign{}
\endlastfoot
\texttt{GENERATE} & \texttt{StateGraph} node & \texttt{splruntime.Generate()} & \texttt{await\ generate()} \\
\texttt{SOLVE} & \texttt{StateGraph} node & \texttt{splruntime.KernelExec()} & \texttt{await\ kernelExec()} \\
\texttt{ASSERT\ ...\ OTHERWISE} & conditional edge & \texttt{if\ !ok} branch & \texttt{if\ (!ok)\ throw} \\
\texttt{CALL\ PARALLEL} & parallel nodes & goroutines + \texttt{sync.WaitGroup} & \texttt{Promise.all({[}...{]})} \\
\texttt{RETURN} (\texttt{COMMIT} is deprecated) & \texttt{END} node & \texttt{return} / error & \texttt{return} / \texttt{throw} \\
\end{longtable}

The DODA principle holds at the compilation level: the same declarative specification produces idiomatic, correct code in each target language without any manual adaptation.

\section*{Appendix G: Runtime Implementation Details}\label{appendix-g-runtime-implementation-details}

\subsection*{G.1 Kernel Substrates}\label{g.1-kernel-substrates}

\texttt{spl3/kernel.py} provides two execution substrates for deterministic nodes:

\textbf{\texttt{IPythonKernel}} runs an out-of-process Jupyter kernel via \texttt{jupyter\_client.KernelManager}. It is lazy-started on the first \texttt{SOLVE} or \texttt{ASSERT} encountered during a workflow run, then held alive for the run's duration. All state --- imported modules, defined variables, intermediate results --- persists across steps within a session. The kernel exposes a thread-locked \texttt{execute()} method that captures stdout and the \texttt{text/plain} repr of the last expression. Python-level errors raise \texttt{KernelExecutionError}, which the executor maps to the SPL exception hierarchy.

\textbf{\texttt{KernelSession}} is a lightweight in-process \texttt{exec()} substrate with a persistent namespace dict. It is used for \texttt{CREATE\ TOOL\_API} body execution, where the overhead of an out-of-process kernel is unnecessary. Two isolation scopes are supported: \texttt{"session"} (shared namespace across the run) and \texttt{"workflow"} (fresh namespace per workflow invocation).

The kernel rung (SymPy / SageMath / Lean 4) is selected via \texttt{-\/-kernel-name} at invocation time; setup instructions for each rung are in Appendix C.

\subsection*{G.2 Template Resolution and Kernel Dispatch Harness}\label{g.2-template-resolution-and-kernel-dispatch-harness}

Before a \texttt{SOLVE} expression is sent to the kernel, the executor resolves \texttt{\{@var\}} interpolations:

\begin{verbatim}
def _resolve_python_template(self, expr: str, state: WorkflowState) -> str:
    return re.sub(
        r'\{@(\w+)\}',
        lambda m: str(state.get_var(m.group(1))),
        expr
    )
\end{verbatim}

The resolved expression is then wrapped in a standard harness before dispatch:

\begin{itemize}
\tightlist
\item
  \textbf{SOLVE:} \texttt{\_spl\_solve\_result\ =\ \{expr\};\ print(str(\_spl\_solve\_result))} --- printed output is captured and assigned to the target \texttt{@var}.
\item
  \textbf{ASSERT:} \texttt{\_spl\_assert\_result\ =\ bool(\{expr\});\ print(\_spl\_assert\_result)} --- the executor checks the output equals \texttt{"True"}; on mismatch the \texttt{OTHERWISE} body executes.
\end{itemize}

This wrapping is invisible to the workflow author: \texttt{SOLVE} and \texttt{ASSERT} behave as if Python expressions natively return SPL variables.

\subsection*{G.3 Adapter Bootstrap and Momagrid Protocol}\label{g.3-adapter-bootstrap-and-momagrid-protocol}

The adapter bootstrap order in \texttt{spl3/adapters/\_\_init\_\_.py} is three-step:

\begin{enumerate}
\def\labelenumi{\arabic{enumi}.}
\tightlist
\item
  \textbf{dd-llm bridge} --- if installed, registered first for all cloud model prefixes (Anthropic, OpenAI, Google, Mistral, Cohere, Together, Groq, Fireworks, Perplexity, DeepSeek).
\item
  \textbf{Bespoke fallbacks} --- \texttt{claude\_cli} and \texttt{openrouter} registered as standalone adapters.
\item
  \textbf{Always-available} --- \texttt{ollama} and \texttt{momagrid} registered unconditionally.
\end{enumerate}

The Momagrid adapter submits tasks via \texttt{POST\ /tasks} and polls \texttt{GET\ /tasks/\{id\}}. The hub response carries a nested result object:

{\footnotesize
\begin{verbatim}
{"state": "COMPLETE", "result": {"content": "...", "latency_ms": 342, "agent_name": "worker-07"}}
\end{verbatim}
}

\texttt{agent\_name} is load-bearing for multi-node routing and audit logging; it must not be flattened into the top-level response.

\subsection*{G.4 Ecosystem Tooling}\label{g.4-ecosystem-tooling}

\textbf{\texttt{spl3\ text2spl}} converts a natural-language description to valid \texttt{.spl} source using Code-RAG: the 70+ cookbook recipes are indexed as description/source pairs and retrieved by similarity. A validation loop parses the generated output, feeds parse errors back to the LLM, and iterates until syntactically valid.

\textbf{\texttt{spl3\ vibe}} performs one-shot NL-to-working-code generation: natural language → \texttt{.spl} workflow → runnable Python implementation + README + test data. The \texttt{-\/-out-dir} flag writes all artifacts to a folder; \texttt{-\/-adapter} selects the generating model.

Both tools are available after \texttt{pip\ install\ spl-llm} with no additional configuration.

\subsection*{G.5 Lean 4 Three-Stage Protocol (SPL Source)}\label{g.5-lean-4-three-stage-protocol-spl-source}

The full SPL implementation of the Lean rung referenced in \S\ref{the-verifier-ladder}:

\begin{verbatim}
-- Initialise the Lean REPL (once per session)
CALL run_python("from spl3.lean_bridge import LeanREPL;
                 _spl_lean = LeanREPL.mathlib().start();
                 print('ready')") INTO @lean_status

-- Stage 1: Formalize
GENERATE formalize_claim(@problem) INTO @lean_stmt
@check_code := f"print(_spl_lean.statement_ok(r'''{@lean_stmt}'''))"
CALL run_python(@check_code) INTO @stmt_ok

-- Stage 2: Typecheck with self-repair loop
@tries := 0
WHILE @tries < @max_tries DO
  EVALUATE @stmt_ok
    WHEN = "False" THEN
      GENERATE fix_formalization(@problem, @lean_stmt, @lean_feedback) INTO @lean_stmt
      CALL run_python(@check_code) INTO @stmt_ok
    ELSE
      @tries := @max_tries
  END
  @tries := @tries + 1
END

-- Stage 3: Prove with self-repair loop
GENERATE write_proof(@lean_stmt) INTO @lean_proof
@prove_code := f"print(_spl_lean.proof_ok(r'''{@lean_stmt}''', r'''{@lean_proof}'''))"
CALL run_python(@prove_code) INTO @proof_ok

@proof_tries := 0
WHILE @proof_tries < @max_tries DO
  EVALUATE @proof_ok
    WHEN = "False" THEN
      GENERATE fix_proof(@lean_stmt, @lean_proof, @lean_feedback) INTO @lean_proof
      CALL run_python(@prove_code) INTO @proof_ok
    ELSE
      @proof_tries := @max_tries
  END
  @proof_tries := @proof_tries + 1
END

RETURN @lean_proof WITH status = 'machine_proved'
\end{verbatim}

\section*{Appendix H: Model Selection for Two-Mode Workflows}\label{appendix-h-model-selection-for-two-mode-workflows}

Two task types arise in a declarative two-mode workflow; they suit different model behaviors.

\begin{longtable}[]{@{}
  >{\raggedright\arraybackslash}p{2.0cm}
  >{\raggedright\arraybackslash}p{5.6cm}
  >{\raggedright\arraybackslash}p{5.5cm}@{}}
\toprule\noalign{}
\begin{minipage}[b]{\linewidth}\raggedright
Property
\end{minipage} & \begin{minipage}[b]{\linewidth}\raggedright
Probabilistic
\end{minipage} & \begin{minipage}[b]{\linewidth}\raggedright
Deterministic
\end{minipage} \\
\midrule\noalign{}
\endhead
\bottomrule\noalign{}
\endlastfoot
Goal & Pattern recognition, format mapping & Exact, reproducible derivation \\
Output & Approximate, context-shaped & Verifiable, same answer every run \\
Extended thinking & Liability --- wastes tokens on work the kernel will do & N/A --- correctness guaranteed by construction \\
Key metric & Structured output compliance & Machine-checkable correctness \\
Speed & Variable & Can be faster than probabilistic for structured problems \\
Role in SPL & LLM (GENERATE, EVALUATE) & Kernel (SOLVE, ASSERT) \\
\end{longtable}

\subsection*{Plan-and-Explain Pattern for Probabilistic Tasks}\label{plan-and-explain-pattern-for-probabilistic-tasks}

The two-mode architecture enables a reusable pattern for workflows that combine LLM reasoning with deterministic verification. The \textbf{plan-and-explain} pattern decomposes the LLM's work into three stages:

\begin{enumerate}
\def\labelenumi{\arabic{enumi}.}
\item
  \textbf{Plan} (probabilistic): The LLM breaks down the problem into a structured format the deterministic engine understands --- e.g., decomposing a math problem into \texttt{expression\textbar{}operation} steps, or translating a natural-language claim into a formal Lean statement.
\item
  \textbf{Execute} (deterministic): The kernel (SymPy, SageMath, Lean, or any Python-callable verifier) processes the plan step-by-step, producing exact, reproducible results. Errors are caught and reported explicitly via exception handling or status codes.
\item
  \textbf{Explain} (probabilistic): The LLM narrates the kernel's output in natural language, providing context, interpretation, or summary. Since the result is already verified, the LLM's role is presentation, not reasoning --- it can use smaller, faster models. This stage is critical for accessibility: mathematicians and domain experts often express their work in cryptic or arcane notation (specialized symbols, terse formalism, rare terminology) that few understand. The explain step translates verified results into accessible prose, making specialized knowledge available to a broader audience --- turning rare textbooks and dense proofs into comprehensible narratives without sacrificing rigor or correctness.
\end{enumerate}

This pattern appears throughout the cookbook: recipe \#67 (symbolic math) plans a chain of algebraic steps, verifies with SymPy, then explains the result; recipe \#76 (Lean proof) formalizes a claim, checks it with the kernel, and interprets the proof badge. The pattern generalizes: any domain with a Python-callable verifier (unit testing, schema validation, graph properties, constraint satisfaction) can use plan-and-explain to combine LLM fluency with deterministic correctness.

In the SPL neurosymbolic experiment the LLM's job is \textbf{probabilistic}: recognize the problem type and map it to \texttt{expr\textbar{}op} format (plan stage). The kernel's job is \textbf{deterministic}: execute each step exactly (execute stage). The LLM then explains the verified result. Thinking-mode models (qwen3.5:9b, deepseek-r1) apply deterministic-style deliberation to the plan stage, exhausting their token budget before emitting any structured output --- a mismatch that plan-and-explain avoids by making the plan stage a format-translation task, not a reasoning task.

The broader implication: declarative composition lowers the LLM capability bar for math workflows. Without SPL, solving calculus requires a model that reasons deterministically. With SPL and the plan-and-explain pattern, it requires only a probabilistic translator --- the kernel handles the derivation, and a smaller model handles the explanation.

\end{document}